\documentclass[manuscript]{aastex62}
\usepackage{natbib}
\usepackage{subfigure}

\received{}
\revised{}
\accepted{}

\submitjournal{ApJ}

\shorttitle{Temporal study of Ton 599}
\shortauthors{Patel et al.}

\begin{document}

\title{Temporal variability and estimation of jet parameters for Ton 599}

\correspondingauthor{S. R. Patel}
\email{sonal.patel@tifr.res.in, sonalpatel.982@gmail.com}

\author{S. R. Patel}
\affiliation{Department of High Energy Physics, Tata Institute of Fundamental Research, Homi bhabha road, Mumbai-400005, India}
\affiliation{Department of Physics, University of Mumbai, Santacruz (E), Mumbai-400098, India}
\author{V. R. Chitnis}
\affiliation{Department of High Energy Physics, Tata Institute of Fundamental Research, Homi bhabha road, Mumbai-400005, India}
\author{A. Shukla}
\affiliation{Institute for Theoretical physics and Astrophysics, Universit$\ddot{a}$t W$\ddot{u}$rzburg, 97074 W$\ddot{u}$rzburg, Germany}
\author{A. R. Rao}
\affiliation{Department of Astronomy and Astrophysics, Tata Institute of Fundamental Research, Homi bhabha road, Mumbai-400005, India}
\author{B. J. Nagare}
\affiliation{Department of Physics, University of Mumbai, Santacruz (E), Mumbai-400098, India}

\begin{abstract}

The TeV blazar Ton 599 has exhibited a peculiar flare in 2017 November. 
The temporal variation of the source is studied using simultaneous $\gamma$-ray
data from $\textit{Fermi}$ Large Area Telescope and radio data 
from Owens Valley Radio Observatory's 40 m telescope, 
over the period of nine years.
Four major flaring periods are observed in the $\gamma$-ray energy band of 
0.1-300 GeV. These periods are studied on a shorter timescale and 
modeled with a time-dependent function containing exponential rising and 
decaying components. The physical parameters of the jet are estimated 
numerically and compared with those reported in the literature. 
During the fourth flare a bunch of high energy photons  ($>$~10~GeV) were 
detected. The two highest energy photons having an energy of 76.9 GeV 
and 61.9 GeV are detected on MJD 58059.0 and 58073.3, respectively.
This observation possibly constrains the $\gamma$-ray emission region to
lie near outer edge or outside the broad line region of size $\sim$0.08~pc. 
The variation of equivalent width of a Mg-II line 
is studied using the spectroscopic data from Steward observatory.
It was observed that the equivalent width of the line varies inversely 
with the underlying power-law continuum.

\end{abstract}


\section{Introduction} \label{sec:intro}
Blazars are a subclass of active galactic nuclei (AGNs), characterized by a
significant variability at all wavelengths and with the flux variation  
timescale ranging from minutes to years. The variability carries plenty of
information about the structure and energy changes happening at the 
center of the source. The emission from blazars is generally thought 
to be coming from the relativistic jets which are beamed towards the 
observer \citep{urry1995}. Blazars are classified into two categories 
based on the observation of the optical spectra, as BL Lacertae (BL Lac) 
and flat spectrum radio quasars (FSRQ). The BL Lac has a featureless 
optical spectrum while the FSRQ shows characteristic strong emission lines 
in an optical spectrum.

The FSRQ, Ton 599 \cite[z=0.725,][]{Schneider2010, Hewett2010}
shows significant variability
across the entire electromagnetic spectrum from radio to $\gamma$-rays. 
Owing to the wide range of variability, Ton 599 has been studied since more 
than four decades in several energy bands. At $\gamma$-ray energies, 
it was detected in the second EGRET catalog \citep{Thompson1995}. Due to better
sensitivity of Large Area Telescope on board $\textit{Fermi}$ 
satellite ($\textit{Fermi}$-LAT), this source was detected in the first three months of 
operation \citep{Abdo1FGL}. 
During the period 1992 to 1996, three strong flares 
having flux of $\sim$3.9 $\times$ 10$^{-10}$ Jy ($\sim$9.4 10$^{-11}$ erg cm$^{-2}$ s$^{-1}$) 
were detected at energies $>$ 100 MeV by EGRET \citep{Thompson1995,Mukherjee1997,Hartman1999}.


The blazars are known to show aperiodic variability. Also
claims of periodicity in radio band have been reported for 
several blazars, for example, A0 0235+16 \citep{Raiteri2001, Liu2006a}, 
PKS 1510-089 \citep{Xie2008}, NRAO~530 \citep{An2013}, OJ~287 
\citep{Sillanpaa1988, Valtaoja2000, Cohen2017} and
PKS 0219-164 \citep{Bhatta2017}.
Ton 599 was regularly monitored by University of Michigan Radio Astronomy
Observatory \cite[UMRAO,][]{Aller1985}. The weekly sampled data collected 
over nearly 33 years upto 2012, were used to search for a characteristic 
periodic variability of the source \citep{Liu2014}. This object also
exhibited a possible quasi periodic oscillation 
having period of $\sim$2.3$\pm$0.1 years in the radio band \citep{Wang2014, Liu2014}. 
The data collected between 2010$-$2013 from RATAN$-$600 radio telescope 
(Special Astrophysical Observatory, Russian Academy of Sciences, 
\citet{Gorshkov1981,Gorshkov2003}) 
and 32 m Zelenchuk and Badary radio telescope \citep{Gorshkov2009}, 
were used to search for intra day variability (IDV). The IDV timescale of 6 h 
at 4.85 GHz was detected on November 10-11, 2012 at the Badary station 
\citep{Gorshkov2014}. The flux density variation of 40 $\%$ 
on timescale of 2.7 h was observed during 2007 February 5 
by the MOJAVE survey program \citep{Lister2005} 
with the VLBA at 15 GHz, 
which had placed this source in between classical IDV and intra-hour 
variable sources \citep{Savolainen2008}.  

Ton 599 is classified as an optically violent variable and a highly 
polarized quasar \citep{Wills1983,Fan2006}.
A long-term study of this source in the optical bands (Johnson V, Cousins 
RI passbands) was carried out for the data collected from 
1997 April to 2002 March, using 1.56 m telescope of the 
Shanghai Astronomical Observatory 
at Sheshan, China, by \citet{Fan2006}. They searched for a possible 
periodicity in the lightcurve and the periods of 1.58 and 3.55 
years were reported. The periodic variation 
consisting of two outbursts having one-year-separating double-peaked 
structure with a gap of 3.2 years in the infrared (JHK band) was also 
reported by \citet{Fan1999} using 
the historic data compiled from literature.

The connection between the $\gamma$-ray and radio emissions has also been
studied for this source and constraints are put on $\gamma$-ray emitting
region in parsec-scale jet by \citet{Ramakrishnan2014}. They interpreted,
using radio data, the identification of four moving components and one stationary
component as the complex change in the jet structure due to formation of
a trailing shock in an inner region of the jet. A radiative transfer 
model \citep{Hughes2011}, was used for this source in order to reproduce 
the observed lightcurve \citep{Aller2014} and the linear polarization.
The helical jet model was also used for this source to explain a 
"double-peak" structure of the total flux density lightcurve 
by \citet{Hong2004} in radio band. 
They infer that the double-peak structure should appear, if the jet 
components always travel along the helical path aligned with the line of sight 
of the observer.

In the present work, a long-term temporal study of the FSRQ Ton 599 in 
$\gamma$-ray energy band is carried out and the jet parameters are estimated.
We identified four major outbursts in this energy band since 2008.
The source is then studied in detail at four epochs during which outbursts
occurred. Also the possible implication of the time lag between 
$\gamma$-ray and radio emission is discussed. The variation of the Mg-II line 
with underlying continuum is studied to see the effect of different flux 
states on the broad line region (BLR). The observations and analysis of the 
data used in this work is discussed in Section~\ref{analysis} and 
results of the analysis are mentioned in Section~\ref{result}. 
In Section~\ref{SED}, the 
spectral energy distributions (SEDs) at four different epochs
are discussed and we conclude our work with a discussion in Section~\ref{discussion} 
and a conclusion in Section~\ref{Conclusion}. A flat $\Lambda$-CDM cosmology 
with $H_0$ = 69.6 km s$^{-1}$ Mpc$^{-1}$, $\Omega_m$ = 0.286, and 
$\Omega_{\Lambda}$ = 0.714 is used in this work \citep{Planck2014}.

\section{Observations and data analysis}
\label{analysis}
In this study, the high energy $\gamma$-ray data from $\textit{Fermi}$-LAT
was obtained for the period of around nine years. The publicly available 
spectroscopic data from SPOL
and radio data from OVRO
were used in this work. Also X-ray and Optical-UV data were taken 
from X-ray telescope on board $\textit{Swift}$ satellite ($\textit{Swift}$-XRT).
In this section,
we describe these data sets and an analysis procedure used in this work.

\subsection{High energy $\gamma$-ray observations from $\textit{Fermi}$-LAT}

The $\textit{Fermi}$-LAT \citep{Atwood2009} is a 
pair-conversion $\gamma$-ray telescope 
operational since  2008. It has a large field of view of 2.4 $sr$, because 
of which it scans the whole sky in $\sim$3 hours period. The Science 
Tools version v10r0p5 and user contributed $\textit{enrico}$ 
software \citep{enrico}  were used for analysis of Ton 599 data. 
The $\textit{Fermi}$-LAT data\footnote{https://fermi.gsfc.nasa.gov/ssc/data/}
in the energy range of 0.1~-~300 GeV were collected during 2008 August 22
(MJD 54700) to 2018 May 22 (MJD 58260) period.
For analysis, the events were extracted from the region of interest (ROI)
of 15$^\circ$ centered around the source position (RA=179.883, Dec.=29.2455).
In order to avoid the contamination of the background $\gamma$-rays from
Earth's limb, zenith angle cut of 90$^\circ$ was applied. A filter of
'(DATA\_QUAL$>$0)\&\&(LAT\_CONFIG==1)' was applied to select the good time
intervals. The unbinned likelihood analysis was performed using
$\textit{gtlike}$ \citep{Cash1979,Mattox1996}. Our model contained 66
point sources from 3FGL catalog. The parameters of sources within 5$^\circ$ 
region around the source were left free to vary. Also the parameters of one
bright source J1224.9+2122, $\sim$9$^\circ$ away from Ton 599 were kept 
free. The likelihood analysis was performed iteratively by removing the
source having significance less than 1 $\sigma$ after each fit.
Our source was modeled with a power law spectrum for the lightcurves.
In the analysis the galactic diffuse emission and the isotropic 
background were modeled using "gll\_iem\_v06.fit" and 
"iso\_P8R2\_SOURCE\_V6\_v06.txt", respectively with the
post launch instrument response function (P8R2\_SOURCE\_V6).
The fluxes were obtained with integration time of one day, two days
and ten days depending on the science goals, with detection criterion
such as maximum likelihood test statistics \cite[TS;][]{Mattox1996}
exceeding 9 ($\sim$3 $\sigma$). For fluxes with the detection of TS$<$9,
the upperlimit at 0.95 confidence level were estimated using the
profile likelihood method \citep{Rolke2005}. To obtain the probability 
of photons with energy above 10 GeV, being associated with Ton 599, 
within 0.5$^\circ$, $\textit{gtsrcprob}$ was used. This tool is provided 
with the $\textit{Fermi}$ science tools.

The spectra were obtained during four epochs, when the source was in 
the high flux state and a fitting was performed using maximum 
likelihood analysis. The functions used for fitting are as follows,

\begin{itemize}
\item[1.]  A Power-law (PL), defined as
\begin{equation}
dN(E)/dE = N_{0} (E/E_p)^{-\Gamma},
\label{pl}
\end{equation}
where, $E_p$=439.38 MeV, which is the value of pivot energy given in 
3FGL catalog and $\Gamma$ is the spectral index.
\item[2.] A broken power-law (BPL), defined as
\begin{equation}
dN(E)/dE = N_{0} (E/E_{break})^{-\Gamma_i},
\label{bpl}
\end{equation}
with i = 1 if E $<$ E$_{break}$ and i = 2 if E $>$ E$_{break}$, $\Gamma_1$ and $\Gamma_2$ 
are the spectral indices before and after the break energy, $E_{break}$, respectively.
\item[3.] A log-parabola (LP), defined as
\begin{equation}
dN(E)/dE = N_{0} (E/E_0)^{-(\alpha + \beta log(E/E_0))},
\label{lp}
\end{equation}
where, E$_0$ is the value of the pivot energy (same as PL), $\alpha$ 
and $\beta$ are the spectral index and curvature parameter, 
respectively.
\item[4.] A Power-law with an exponential cutoff (PLEC), defined as
\begin{equation}
dN(E)/dE = N_{0} (E/E_p)^{-\Gamma_{PLEC}} exp(-E/E_c),
\label{plex}
\end{equation}
where E$_p$ is the value of pivot energy (same as PL). $\Gamma_{PLEC}$ 
is the spectral index and $E_c$ is the value of cutoff energy.
$N_0$ is the normalization parameter for all the models.
\end{itemize}

\begin{figure*}[ht!]
\centering
\includegraphics[scale=0.9]{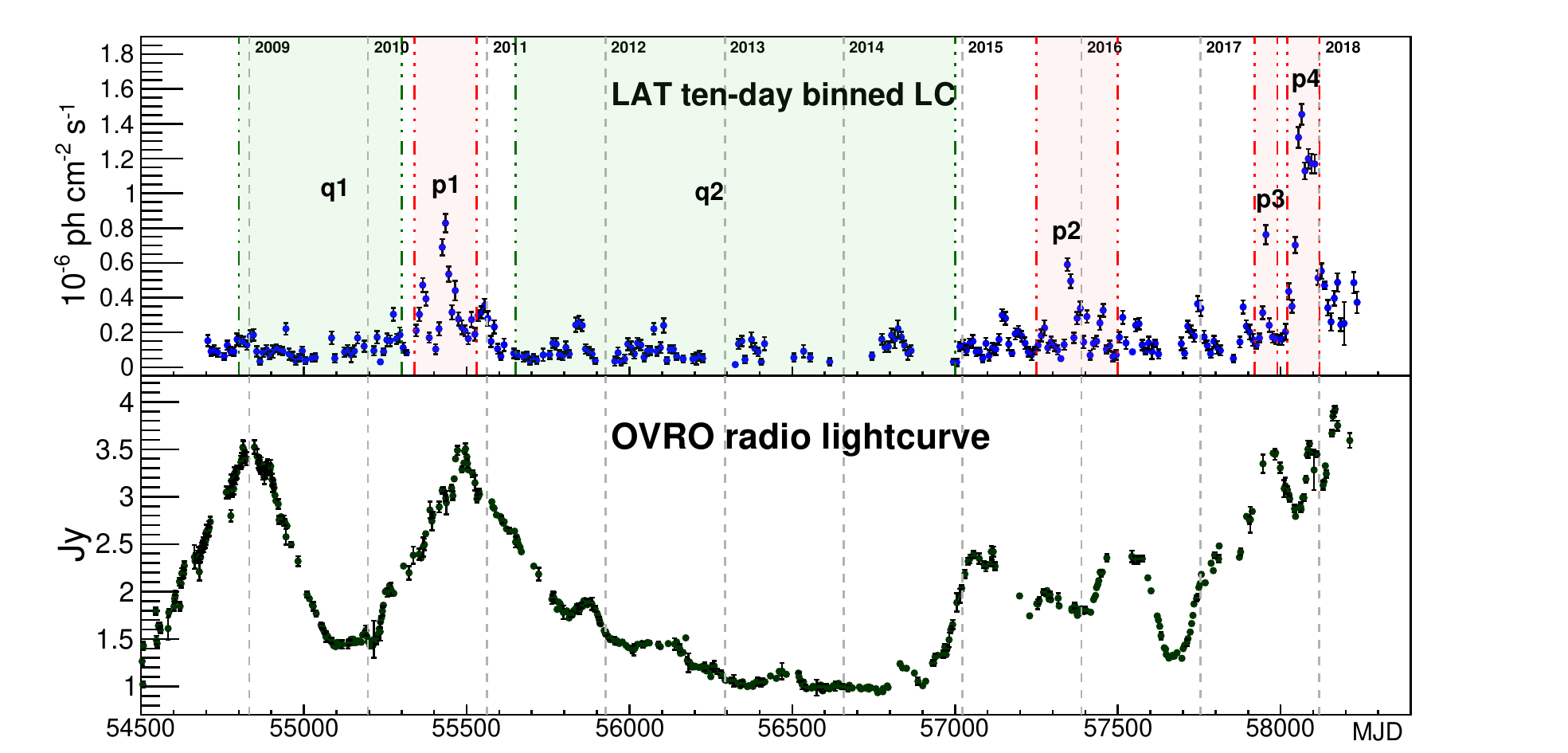}
\caption{Nine years lightcurve of Ton 599 from MJD 54700 - 58260, 
$\textit{Panel-1}$ : Ten days binned 0.1 - 300 GeV flux in 
10$^{-6}$ ph cm$^{-2}$s$^{-1}$, $\textit{Panel-2}$ : Radio flux density 
at 15 GHz in Jy; The shaded regions between two red dot-dashed lines correspond to 
the periods which are studied in detail and marked as p1, 
p2, p3, and p4. The shaded periods between two green dot-dashed lines,
marked as q1 and q2 correspond to quiescent states of the source. 
The gray dashed lines correspond to start of the year.}
\label{LC_10}
\end{figure*}

\subsection{X-ray observations from $\textit{Swift}$-XRT}
\label{xray}

The soft X-ray data\footnote{https://heasarc.gsfc.nasa.gov/docs/archive.html} 
in the energy range of 0.3~-~8 keV were obtained from
$\textit{Swift}$-XRT \citep{xrt} and used in the estimation of 
minimum Doppler factor. The data were analyzed using XRT data 
analysis software (XRTDAS) distributed within the HEASOFT package (v6.19).
The cleaned event files were generated using \textit{xrtpipeline-0.13.2} 
tool, and \textit{xrtproduct-0.4.2} tool was used to generate spectral
files. 
The spectra observed during flares p1 and p4 (marked in Figure~\ref{LC_10}) 
were added using an \textit{addspec} tool and rebinned with 
a minimum of 20 photons per bin. The final spectrum was then fitted with 
an absorbed power-law model with a neutral hydrogen column density, 
$n_H$ fixed at the galactic value of 1.81 $\times$ 10$^{20}$ cm$^{-2}$ 
\citep{nh}. 

\subsection{Optical observations from SPOL}

The spectroscopic data\footnote{http://james.as.arizona.edu/$\sim$psmith/Fermi/} from SPOL CCD Imaging/Spectropolarimeter
at Steward observatory at the University of Arizona were used in this work. 
It combines polarizing optics and a transmission-optics spectrograph
into a self-contained, portable, and high-throughput instrument. The complete
details and explanation of the data is given in \citet{Smith2009}. 
The SPOL regularly monitors Ton 599 as a part of the 
$\textit{Fermi}$ multiwavelength support program. 
The V-band magnitude and the linear polarization data since 2011 
were obtained from SPOL.

\subsection{Optical-UV observations from \textit{Swift}-UVOT}

The optical-UV data\footnote{https://heasarc.gsfc.nasa.gov/docs/archive.html} 
from \textit{Swift}-UVOT \citep{Roming2005} were also 
analyzed and used to estimate the magnetic field in the emission region. 
The optical data are provided in three filters, 
viz. \textit{V, B} and \textit{U}, while the data in UV bands
are available in \textit{UVW1}, \textit{UVM2} and \textit{UVW2} 
filters. The observed magnitudes were corrected for a galactic extinction 
of $E_{B-V}$ = 0.0171 mag \citep{Schlafly2011}. The corrected observed 
magnitudes at all six wavelengths were then converted into the flux using 
zero point magnitudes \citep{Poole2008}.

\subsection{Radio observations from OVRO}
Ton 599 is also regularly monitored in radio by 40~m Owens Valley Radio
Observatory (OVRO) as a part of $\textit{Fermi}$ 
monitoring program, since 2008 January 1 (MJD 54473). 
The complete details of the observing program and
calibration are given in \citet{Richards2011} and \citet{Richards2012}.
The off-axis dual-beam optics and a cryogenic reciver with
3 GHz bandwidth centered at 15 GHz are used in the telescope. 
The total flux density measurements are performed using 'double-switching'
procedure. This method removes receiver gain fluctuations, 
ground and atmospheric pick-up effects. The typical uncertainty in 
total flux density measurement is about 3 mJy. This uncertainty 
includes the contribution from total system temperature
($\sim$55 K at zenith), atmospheric and cosmic microwave background temperature.
Besides the measured error, the 2 \%  error comes from residual pointing.
These two errors are added in quadrature for each measurement.  
All the data\footnote{http://www.astro.caltech.edu/ovroblazars/data.php?page=data\_query}, 
since 2008 January 1, till 2018 April 4 (MJD 58212) are used in this work.


\section{Results}
\label{result}

We have extensively studied the long-term lightcurve in $\gamma$-ray 
band (0.1 - 300 GeV) and identified four periods during which the 
source was in high flux state. 
The flares were selected based on flux exceeding 
5.5 $\times$ 10$^{-7}$ph cm$^{-2}$ s$^{-1}$. 
This flux corresponds to 3 $\sigma$ value of the mean background
flux of 2.36 $\times$ 10$^{-7}$ph cm$^{-2}$ s$^{-1}$. 
The periods around the peak flux were chosen arbitrarily for detailed study.
The detailed activity of the source in
these four periods is discussed in this section.

\subsection{Long-term lightcurve}
The Figure~\ref{LC_10} shows ten days binned $\textit{Fermi}$-LAT 
lightcurve encompassing four high flux states marked by p1, p2, p3, and p4. 
It can be seen that the first high flux state (p1) was observed on MJD 
55435 $\pm$ 5 with the flux of (8.3 $\pm$ 0.4) $\times$ 10$^{-7}$ ph 
cm$^{-2}$ s$^{-1}$. After that the source went into the quiescent state 
for almost five years with average flux level of $\sim$10$^{-7}$ ph 
cm$^{-2}$ s$^{-1}$. Towards the end of 2015 i.e on MJD 57345 $\pm$ 5, 
it again went into the flaring state (p2) during which the observed 
flux was (5.9 $\pm$ 0.4) $\times$ 10$^{-7}$ ph cm$^{-2}$ s$^{-1}$, 
which is $\sim$40~$\%$ lower than p1. The third flare (p3) was observed 
almost after one and half year in the middle of the year 2017 
(MJD 57955 $\pm$ 5). The peak flux during p3, (7.62 $\pm$ 0.54) 
$\times$ 10$^{-7}$ ph cm$^{-2}$ s$^{-1}$, was comparable to that of p1. 
Recently the source had again gone into the high flux state (p4) of 
(14.5 $\pm$ 0.6) $\times$ 10$^{-7}$ ph cm$^{-2}$ s$^{-1}$, which is the 
highest ever detected flux in $\textit{Fermi}$-LAT era for this source.
We have studied these four flares in detail by scanning these periods
on shorter timescale of one day except for p2, where the binning of two 
days is used, due to sparse sampling. Through detailed observations 
we found several sub-flares during major flares, p1, p2, p3, and p4, 
which is discussed in this section along with the flare fitting method. 
The change of polarization in optical band during p4 is also discussed in
this section. Due to sparse sampling of the polarization data during 
p1, p2, and p3, it is not studied for these periods.

\subsubsection{Flare fitting}
\label{FlareFitting}

In order to understand the flaring pattern within the major flare, we fit
each individual component and a constant baseline flux in the whole one/two-day 
binned lightcurve. This decomposition of the flare is relatively simple but 
potentially very useful way of understanding the flux variations in the
various energy bands \citep{Valtoja1999}. 
This method is also very useful in probing the shorter timescale
(of the order of minutes) variability seen at $\gamma$-ray energies.
The time dependent profile used for each component is,

\begin{equation}
F(t) = 2 F_0 (\exp^{(t_0-t)/T_r} + \exp^{(t-t_0)/T_d})^{-1},
\label{EqFlare}
\end{equation}
where, $F(t)$ is the flux at time $t$, $t_0$ is the sub-flare 
peak time, $T_r$ and $T_d$ are the rise and fall time respectively 
and $F_0$ is the flux at time $t_0$, which represents the amplitude of 
the sub-flare.

We first fitted each individual sub-flaring component with
the function given in Equation~\ref{EqFlare} and obtained the function parameters.
Then the final best-fit is obtained for the total function consisting of n such 
sub-flaring components and one constant background component. We  use the
fit parameters of individual sub-flares as the starting parameters while
performing the fit of the total function for a given period. 
The residual of the fit which is the ratio of the difference of the
model and observed flux to the flux-error ((Model flux - observed flux)/flux error)
is given in the second panel of Figures~\ref{lc1},~\ref{lc2},
~\ref{lc3} and ~\ref{lc4}, to validate the combined fit of each period. 
Similar method was used in the analysis of temporal structure of the 
ten bright blazars in the work by \citet{Abdo2010}. For 3C~454.3, it was also used in 
studying an exceptional event of 5 day $\gamma$-ray outburst occurred in 
2010 November \citep{Abdo2011} and for outbursts observed in 2009 December 
and 2010 April \citep{Ackermann2010}.\\ 
We also note that in the combined fit, some data points lie above the 
model fit and they seem to be sub-flaring components but we did not 
model such components as long as its residuals are within 3 $\sigma$.
This could be due to superposition of small amplitude flare over
ongoing large amplitude but slowly rising/decaying flare.

\subsubsection{Flare p1}
\label{flarep1}

\begin{figure*} 
\centering
\includegraphics[scale=0.7]{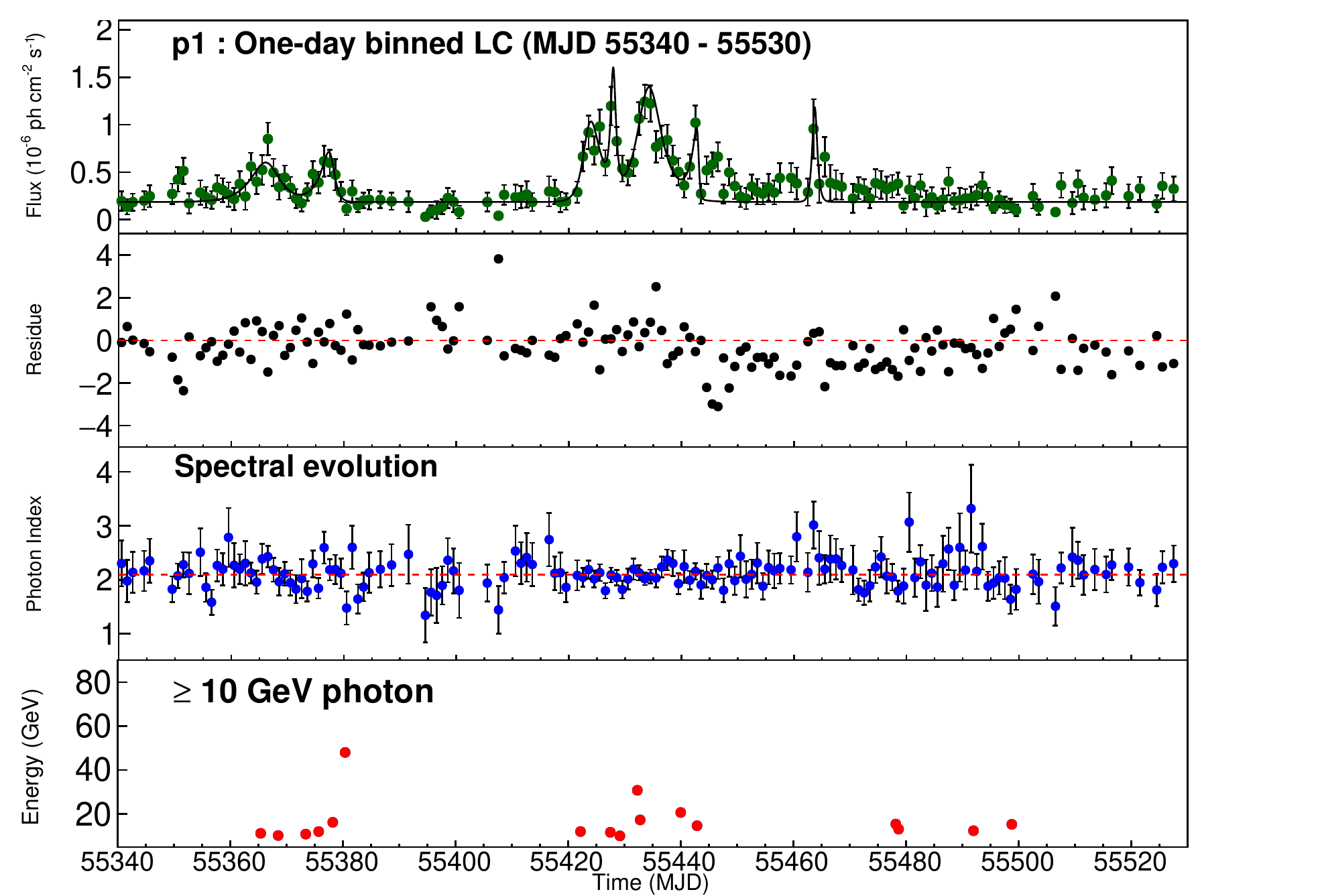}
\caption{\textit{Panel-1}: Lightcurve fitted with seven flaring components 
and constant background component, $\chi^{2}/ndf$~=~1.47;
\textit{Panel-2}: Residual of model fit;
\textit{Panel-3}: Power-law index variation during the flare-p1, 
red dashed line corresponds to the value of spectral index given in 3FGL 
catalog for Ton 599; $\textit{Panel-4}$: Photons with energy 
$\geqslant$~10 GeV associated with the source with 
probability $\geqslant$~0.99.}
\label{lc1}
\end{figure*}

For a detailed study of the flare p1, we generated the one day binned lightcurve
for the period MJD 55340-55530, as shown in the first panel
of Figure-~\ref{lc1}.
In this period we identified seven sub-flares, which were fitted using the
temporal profile as per the Equation~\ref{EqFlare} and the constant baseline
flux. The parameters of the fit are mentioned in Table~\ref{lctab}.
During this flare there was no significant change in the spectral shape,
as can be seen in the third panel of Figure~\ref{lc1}. The highest
energy (HE) photon detected during this period had the energy of 48.02 GeV.
It was detected on MJD 57380, when the second component of the major
flare was decaying.

\subsubsection{Flare p2}
\label{flarep2}

\begin{figure*} 
\centering
\includegraphics[scale=0.7]{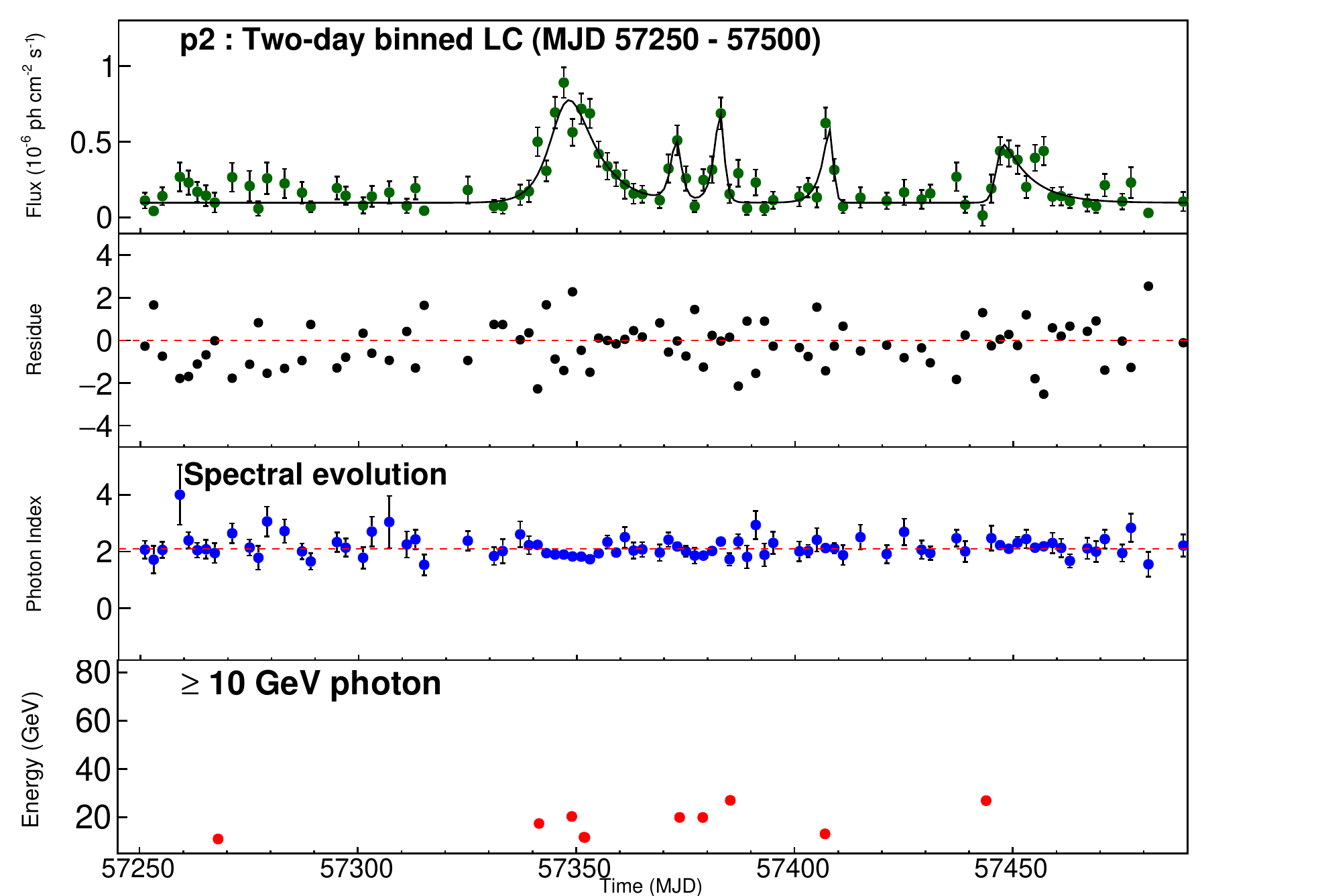}
\caption{\textit{Panel-1}: Lightcurve fitted with five flaring components 
and constant background component, $\chi^{2}/ndf$~=~1.50;
\textit{Panel-2}: Residual of model fit;
\textit{Panel-3}: Power-law index variation during the flare-p2, red 
dashed line corresponds to the value of spectral
index given in 3FGL catalog for Ton 599; $\textit{Panel-4}$: Photons 
with energy $\geqslant$~10 GeV associated with the source 
with probability $\geqslant$~0.99.}
\label{lc2}
\end{figure*}

To understand the sub-structure in the flare p2, we made the two-day binned
lightcurve for MJD 57250-57500. We observe five sub-flares during this period.
The combined fit was obtained by fitting five components to these sub-flares
and one constant background component (see Figure~\ref{lc2}).
The fit parameters are given in Table-~\ref{lctab}. During this period also
spectrum was almost flat and the highest photon energy of 27.02 GeV was
detected during this period.

\subsubsection{Flare p3}
\label{flarep3}

\begin{figure*} 
\centering
\includegraphics[scale=0.7]{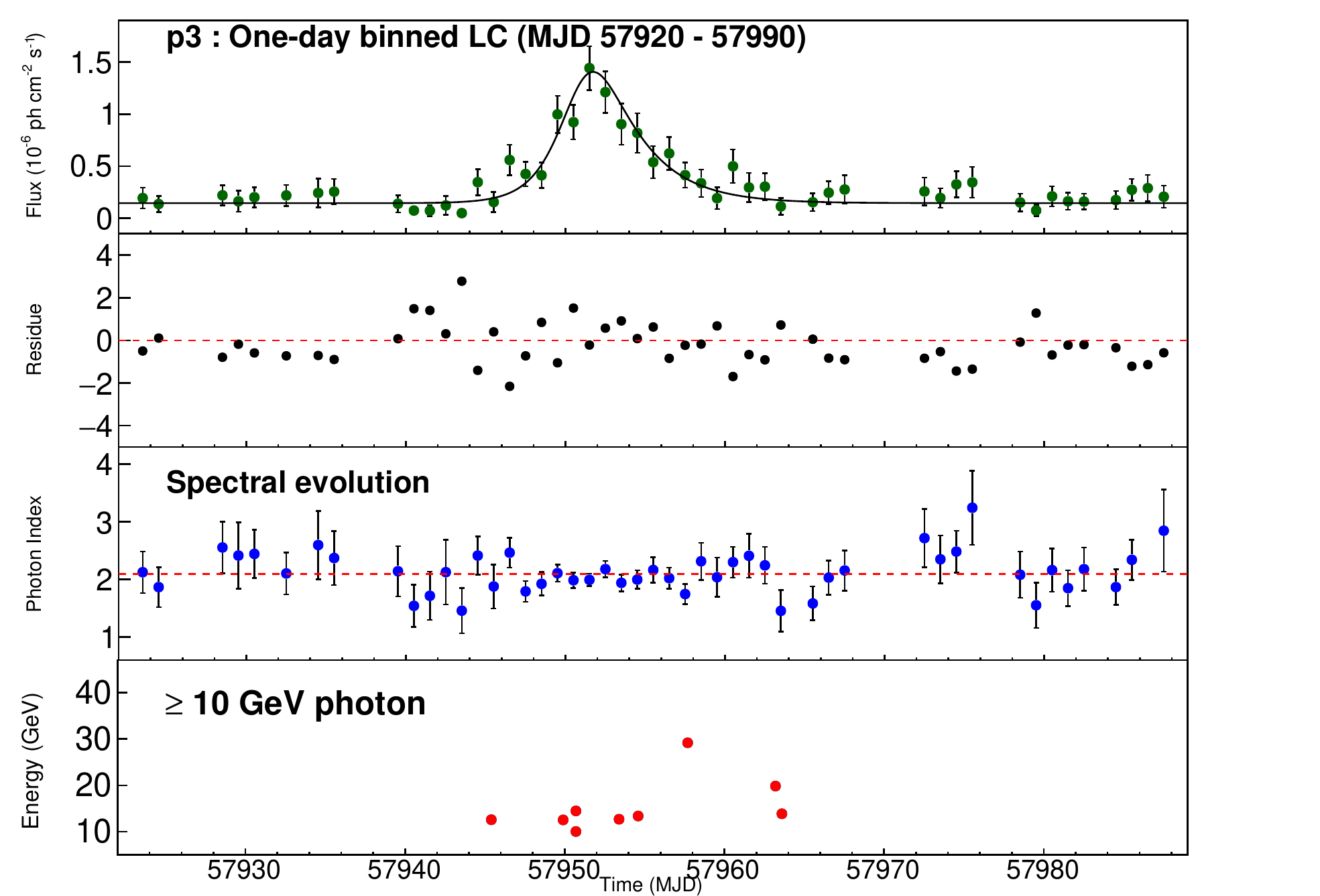}
\caption{\textit{Panel-1}: Lightcurve fitted with one flaring component 
and constant background component, $\chi^{2}/ndf$ = 1.00;
\textit{Panel-2}: Residual of model fit;
\textit{Panel-3}: Power-law index variation during the flare-p3, 
red dashed line corresponds to the value of spectral
index given in 3FGL catalog for Ton 599; $\textit{Panel-4}$: 
Photons with energy $\geqslant$ 10 GeV associated with the source 
with probability $\geqslant$~0.99. 
}
\label{lc3}
\end{figure*}

One-day binned lightcurve for MJD 57920-57990 was generated to
analyze the flare-p3. During p3 one long duration sub-flare was resolved.
This lightcurve was fitted with one flaring component and one constant
background component. The fit is shown in Figure-\ref{lc3} and corresponding
fit parameters are mentioned in Table~\ref{lctab}. The flat spectral evolution
was consistent with p1 and p2, which can be seen in Figure-~\ref{lc3}.
Only few HE photons were detected during p3, with the highest one of
energy 28.9 GeV on MJD 57957.

\subsubsection{Flare p4}
\label{flarep4}

\begin{figure*}[!ht]
\centering
\includegraphics[scale=0.9]{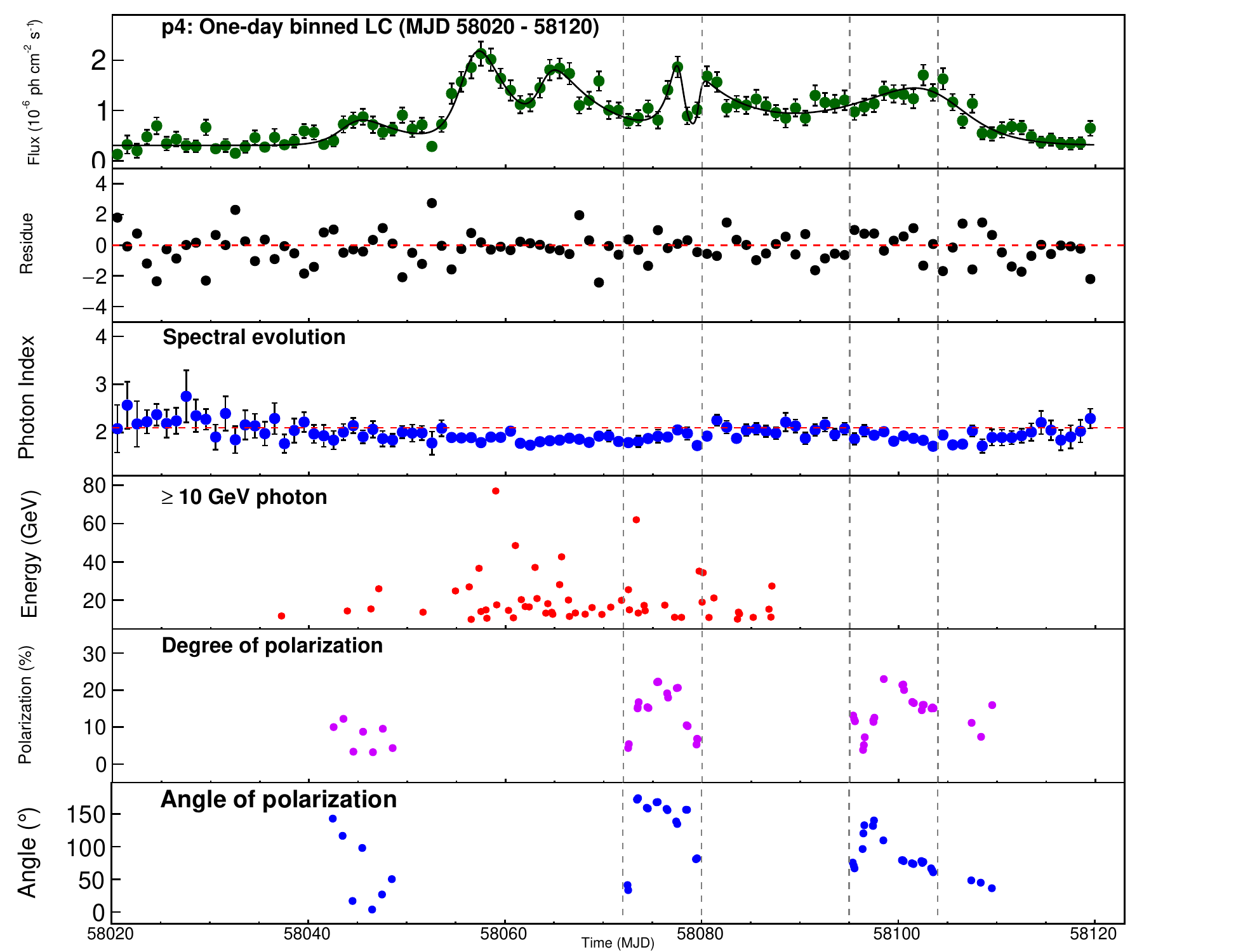}
\caption{\textit{Panel-1}: Lightcurve fitted with six flaring components
and a constant background component, $\chi^{2}/ndf$ = 1.32;
\textit{Panel-2}: Residual of model fit;
\textit{Panel-3}: Power-law index variation during the flare-p4, red
dashed line corresponds to the value of spectral index given in 3FGL
catalog for Ton 599;
\textit{Panel-4}: Photons with energy $\geqslant$ 10 GeV
associated with the source with probability $\geqslant$ 0.99.
\textit{Panel-5}: The degree of polarization in percentage and
\textit{Panel-6}: The angle of polarization in degrees from SPOL data.
The periods during which change in degree of polarization of $\gtrsim$70~$\%$
was observed are shown by dotted gray lines.
}
\label{lc4}
\end{figure*}

The peculiar flare p4 is the brightest among the four flares observed in 
$\gamma$-ray band. This flare was studied by binning the lightcurve for 
MJD 58020 - 58120 in one day bin. Six sub-flares were found during this 
flare and fitted with six flaring components plus one background component. 
The combined fit is shown in Figure~\ref{lc4}.
The parameters of this combined fit are mentioned in Table~\ref{lctab}.
The third panel in Figure~\ref{lc4} shows the variation of the spectral index
during p4, which is consistently harder than previous three flares. Also
this flare was different in a sense that the bunch of HE photons were detected
during this period, with the highest photon energy of 76.94 GeV on MJD 58059.

The variation in the degree of polarization and the 
angle of polarization was also seen in optical band during flare-p4.
These variations are shown in panel-5 and panel-6 of Figure~\ref{lc4},
respectively.
The degree of polarization was found to increase from 
(4.35 $\pm$ 0.05)$\%$ to (22.3 $\pm$ 0.1)$\%$ and again decrease to 
(6.72 $\pm$ 0.06)$\%$ in period of 8 days from MJD 58072 to 58079.
Similar variation was seen during the next observation season, where
it was observed to increase from (3.83 $\pm$ 0.1)$\%$ to (22.98 $\pm$ 0.17)$\%$
and again decrease to (7.37 $\pm$ 0.04)$\%$ in period MJD 58096-58108.
These periods are shown by dashed lines in last two panels of Figure~\ref{lc4}.
In relatively low flux state, the degree of polarization was seen to be
almost constant. \citet{Wills1983} observed the high degree of linear
polarization of 28.05 $\%$, for this source, on 1981 January 31.
The model proposed by \citet{Marscher2008} suggests that the optical polarization 
can increase substantially during first flare when a disturbance
passes through acceleration and collimation zone. 
However, this does not happen for the second flare during 
which radio emissions take place and the disturbance crosses a 
standing shock region. The comparision of radio and optical
polarization data would be useful in better understanding the $\gamma$-ray
flares. Due to sparse sampling of optical polarization
data in our case, we could not relate radio and optical polarization data.


\begin{table*}
\centering
\caption{Parameters of components identified during flare-p1,p2,p3 and p4 
as shown in Figure~\ref{lc1},~\ref{lc2},~\ref{lc3} and ~\ref{lc4}, 
respectively}
\begin{tabular}{ccccc}
\hline
\hline
Sub flare & $t_0$  & $F_0$  & $T_r$  & $T_d$ \\
          & MJD & 10$^{-7}$ph${^{-1}}$ cm$^{-2}$ s$^{-1}$ & day & day \\
\hline
p1-c1 & 55366.7$\pm$1.0 & 4.13$\pm$ 0.84 & 2.54 $\pm$ 1.08 & 1.97 $\pm$0.88 \\
p1-c2 & 55377.8$\pm$1.2 & 4.67$\pm$ 1.46& 1.71$\pm$0.81 & 0.61$\pm$0.31 \\
p1-c3 & 55423.4$\pm$0.5 & 0.80$\pm$ 0.00 & 0.95$\pm$0.32 & 1.86$\pm$0.42 \\
p1-c4 & 55427.9$\pm$0.2 & 12.00$\pm$0.00 & 0.40$\pm$0.19 & 0.40$\pm$0.15 \\
p1-c5 & 55434.1$\pm$0.4 & 12.00$\pm$0.00 & 1.74$\pm$0.31 & 2.11$\pm$0.39 \\
p1-c6 & 55443.0$\pm$0.3 & 5.45$\pm$1.41 & 1.22$\pm$0.55  & \textbf{0.17$\pm$0.14} \\
p1-c7 & 55463.7$\pm$0.6 & 10.00$\pm$0.00 & 0.40$\pm$0.28 & 0.40$\pm$0.40 \\
\hline
\hline
p2-c1 & 57346.8$\pm$1.51 & 6.42$\pm$ 0.64 & 3.11$\pm$0.90 & 6.09$\pm$1.23 \\
p2-c2 & 57372.8$\pm$1.95 & 4.00$\pm$ 0.00 & 1.11$\pm$0.94 & 0.98$\pm$0.75 \\
p2-c3 & 57383.1$\pm$0.95 & 5.64$\pm$ 1.72 & 1.34$\pm$0.72 & 0.67$\pm$0.39 \\
p2-c4 & 57408.4$\pm$0.59 & 4.00$\pm$ 0.00 & 1.96$\pm$0.61 & 0.49$\pm$0.39 \\
p2-c5 & 57446.4$\pm$0.78 & 2.66$\pm$ 0.56 & 0.73$\pm$0.41 & 6.45$\pm$1.60 \\
\hline
\hline
p3-c1 & 57951.1$\pm$0.4 & 12.00$\pm$ 0.00 & 1.45$\pm$0.25 & 2.78$\pm$0.37 \\
\hline
\hline
p4-c1 & 58044.2$\pm$0.79 & 4.30 $\pm$ 0.66 & 1.38 $\pm$ 0.80 & 4.47$\pm$1.00 \\
p4-c2 & 58056.4$\pm$0.25 & 16.23$\pm$1.07 & 1.23$\pm$0.16 & 3.37$\pm$0.36 \\
p4-c3 & 58063.8$\pm$0.43 & 8.74$\pm$0.96 & 0.82$\pm$0.44 & 6.06$\pm$0.61 \\
p4-c4 & 58077.8$\pm$0.21 & 11.52$\pm$1.43 & 1.11$\pm$0.29 & \textbf{0.42$\pm$0.13} \\
p4-c5 & 58079.7$\pm$0.22 & 5.80$\pm$0.91 & 0.21$\pm$0.12 & 5.76$\pm$0.77 \\
p4-c6 & 58104.4$\pm$0.70 & 9.62$\pm$0.57 & 9.51$\pm$0.84 & 2.99$\pm$0.49 \\
\hline
\end{tabular}
\label{lctab}
\tablecomments{A constant baseline flux values of (1.85 $\pm$ 0.08)
$\times$10$^{-7}$, (9.78 $\pm$ 0.87)$\times$10$^{-8}$,
(1.44 $\pm$ 0.15)$\times$10$^{-7}$ and (3.07 $\pm$ 0.19)$\times$10$^{-7}$ ph cm$^{-2}$ s$^{-1}$ 
for flare-p1,p2,p3 and p4, respectively were also fitted to the data along 
with sub-flaring components. The decay times shown in boldface 
were used in calculation of minimum Doppler factor.}
\end{table*}

\subsubsection{Doubling/halving time}
\label{doublingtime}
\begin{table*}
\centering
\caption{Doubling/halving timescales for lightcurves shown in 
Figures~\ref{lc1},~\ref{lc2},~\ref{lc3} and ~\ref{lc4}}
\resizebox{\textwidth}{!}{
\begin{tabular}{cccccccc}
\hline
\hline
Flare & $t_1$ & $t_2$ & Flux-1 & Flux-2 & Doubling/halving time & significance & Rise/decay \\
      & (MJD) & (MJD) & ($10^{-7}$ph cm$^{-2}$s$^{-1}$)  & ($10^{-7}$ph cm$^{-2}$ s$^{-1}$)  & (day)  & ($\sigma$) & (R/D)\\
\hline
   &55407.5 & 55408.5 &0.39$\pm$0.38&  2.61$\pm$1.04&  0.37&   5.82& R\\
p1-c3   &55421.5 & 55422.5 &2.88$\pm$1.11&  6.65$\pm$1.54&  0.83&   3.40& R\\
p1-c4   &55426.5 & 55427.5 &5.98$\pm$1.34&  11.98$\pm$2.03& 0.99&   4.46& R\\
p1-c5   &55431.5 & 55432.5 &5.99$\pm$1.39&  10.64$\pm$1.72& 1.21&   3.32 & R\\
p1-c6   &55441.5 & 55442.5 &5.58$\pm$1.28&  10.22$\pm$1.80& 1.44&   3.63 & R\\
p1-c6   &55442.5 & 55443.5 &10.22$\pm$1.80& 2.73$\pm$1.07&  0.52&   4.17 & D\\
p1-c7   &55462.5 & 55463.5 &2.87$\pm$1.43&  9.57$\pm$3.14&  0.58&   4.69 & R\\
   &55506.5 & 55507.5 &0.79$\pm$0.51&  3.62$\pm$1.30&  0.46&   5.53 & R\\  
\hline
\hline
   &57277.0& 57279.0 &0.59$\pm$0.47& 2.59$\pm$1.04& 0.94 & 4.28 & R\\
   &57339.0& 57341.0 &1.73$\pm$0.72& 5.00$\pm$0.95& 1.31 & 4.55 & R\\
   &57343.0& 57345.0 &3.08$\pm$0.68& 6.93$\pm$1.04& 1.71 & 5.68 & R\\
p2-c1   &57347.0& 57349.0 &8.90$\pm$1.01& 5.63$\pm$0.89& 3.02 & 3.23 & D\\
p2-c2   &57369.0& 57371.0 &1.14$\pm$0.53& 3.23$\pm$0.93& 1.33 & 4.00 & R\\
p2-c3   &57377.0& 57379.0 &0.75$\pm$0.38& 2.48$\pm$0.71& 1.16 & 4.54 & R\\
p2-c3   &57381.0& 57383.0 &3.16$\pm$0.88& 6.87$\pm$1.06& 1.79 & 4.21 & R\\
p2-c4   &57383.0& 57385.0 &6.87$\pm$1.06& 1.56$\pm$0.61& 0.94 & 5.01 & D\\
p2-c4   &57389.0& 57391.0 &0.61$\pm$0.43& 2.32$\pm$0.86& 1.03 & 3.98 & R\\
   &57405.0& 57407.0 &1.33$\pm$0.67& 6.23$\pm$1.03& 0.89 & 7.29 & R\\
   &57409.0& 57411.0 &3.15$\pm$0.73& 0.73$\pm$0.43& 0.95 & 3.31 & D\\
   &57457.0& 57459.0 &4.39$\pm$0.95& 1.38$\pm$0.60& 1.20 & 3.19 & D\\
   &57469.0& 57471.0 &0.75$\pm$0.43& 2.14$\pm$0.75& 1.32 & 3.25 & R\\
\hline
\hline
   &57943.5& 57944.5 &0.48$\pm$0.39&  3.45$\pm$1.25&  0.35&   7.64& R\\
   &57945.5& 57946.5 &1.54$\pm$0.96&  5.58$\pm$1.46&  0.54&   4.16& R\\
   &57948.5& 57949.5 &4.12$\pm$1.22&  9.97$\pm$1.78&  0.78&   4.79& R\\
p3-c1   &57950.5& 57951.5 &9.22$\pm$1.65&  14.46$\pm$2.10& 1.55&   3.13& R\\
\hline
\hline
   &58028.5& 58029.5 &2.89$\pm$1.15& 6.67$\pm$1.56&   0.83&   3.28& R\\
   &58051.5& 58052.5 &7.18$\pm$1.42& 2.87$\pm$1.09&   0.75&   3.03& D\\
   &58052.5& 58053.5 &2.87$\pm$1.09& 7.26$\pm$1.52&   0.75&   4.03& R\\
p4-c2   &58053.5& 58054.5 &7.26$\pm$1.52& 13.38$\pm$2.02& 1.13&   4.02& R\\
   &58075.5& 58076.5 &8.10$\pm$1.65& 14.09$\pm$1.78& 1.25&   3.63& R\\
p4-c4   &58077.5& 58078.5 &18.65$\pm$2.07& 8.93$\pm$1.70& 0.94&   4.69& D\\
p4-c5   &58079.5& 58080.5 &10.17$\pm$1.5& 16.83$\pm$1.97& 1.38 & 4.40 & R\\
   &58090.5& 58091.5 &8.49$\pm$1.47& 13.00$\pm$2.01& 1.63 & 3.10&R \\
p4-c6   &58107.5& 58108.5 &11.40$\pm$1.79& 5.50$\pm$1.22& 0.95& 3.29& D\\
\hline
\end{tabular}
}
\label{DoublingTab}
\tablecomments{The doubling/halving time is in observed frame.}
\end{table*}
The flux doubling/halving timescale analysis was done with the 
data points of entire period of four shorter timescale lightcurves, 
independent of flare fitting method. This analysis was carried out 
to acquire the confidence on rise and decay times that are obtained 
from the flare fitting method.
We scanned all four shorter timescale lightcurves for the fastest flux
doubling/halving time using following equation,
\begin{equation}
F(t_2) = F(t_1) 2^{(t_2 - t_1)/\tau_d},
\label{DoublingTime}
\end{equation}
where F($t_1$) and F($t_2$) are the fluxes at time $t_1$ and $t_2$, 
respectively, and $\tau_d$ is the flux doubling/halving timescale. 
The doubling/halving timescale from consecutive time instances,
having significance above 3~$\sigma$ are listed in the Table~\ref{DoublingTab}. 
The fastest doubling/halving timescales 
obtained using this method during four flares are 0.37, 0.89, 0.35 
and 0.75 days with the significance of 5.82~$\sigma$, 7.29~$\sigma$, 
7.64~$\sigma$, and 4.03~$\sigma$, respectively.
We have checked doubling/halving time for its
coincidence with sub-flaring components and it is mentioned in 
the first column of Table~\ref{DoublingTime}.
It is seen that the most of the sub-flaring components during 
all four flares coincide with the significant doubling/halving time.
Also, the flux doubling/halving timescales are consistent with the 
corresponding rise/decay times obtained from the flare fitting method.

\subsection{Variation of Mg-II line}

\begin{figure}[!h]
\centering
\includegraphics[scale=0.45]{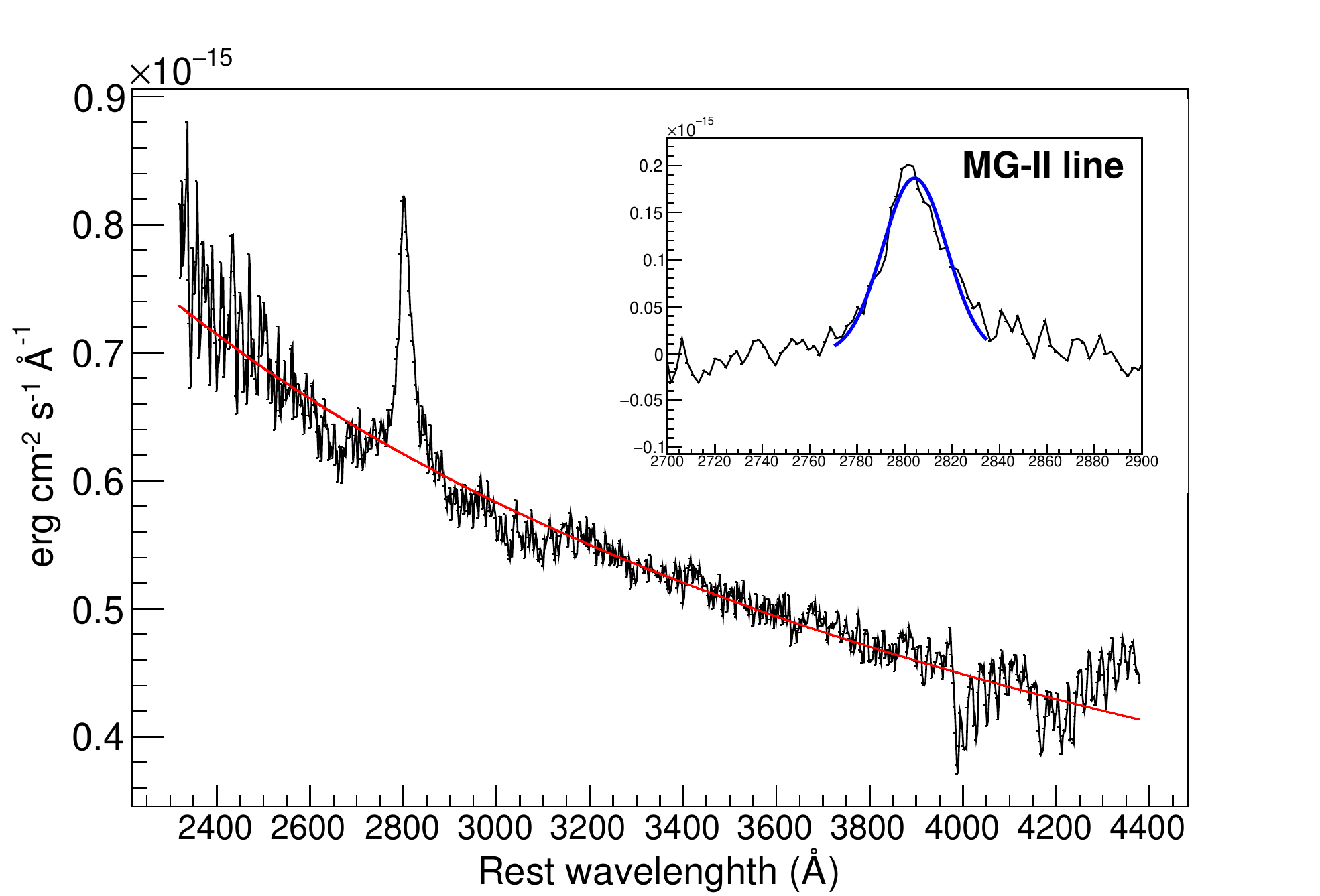}
\caption{An optical spectrum from SPOL with continuum fitted with a 
power-law which is shown by red curve; Inset : A Gaussian fit to the Mg-II 
line after subtracting the power law continuum.}
\label{mg}
\end{figure}

To study the spectroscopic behavior of Ton 599, optical spectra were
obtained from SPOL observatory. These spectra show the prominent 
Mg-II line at wavelength of 2800 {\AA}, which is red-shifted to 
4830 {\AA} for Ton 599. One of the spectra for
Ton 599 from SPOL observations is shown in Figure~\ref{mg}.
For this source we have obtained 140 spectra during the period of 2011 to 2018,
to study the variation of the Mg-II line parameters.

\begin{figure}
\centering
\includegraphics[scale=0.45]{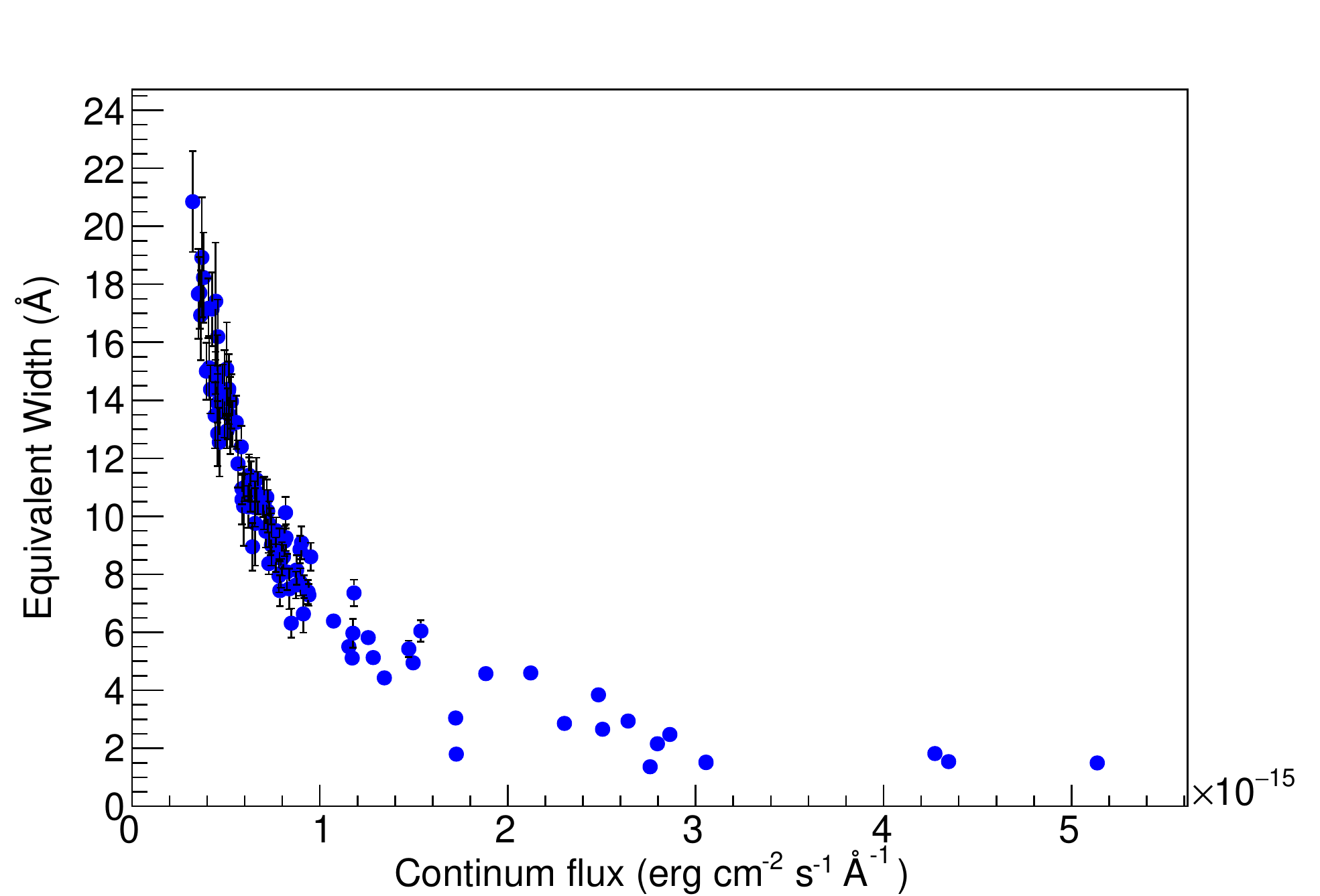}
\caption{The variation of the equivalent width of the Mg-II line with 
continuum flux}
\label{EW}
\end{figure}

\begin{figure}[!h]
\centering
\includegraphics[scale=0.45]{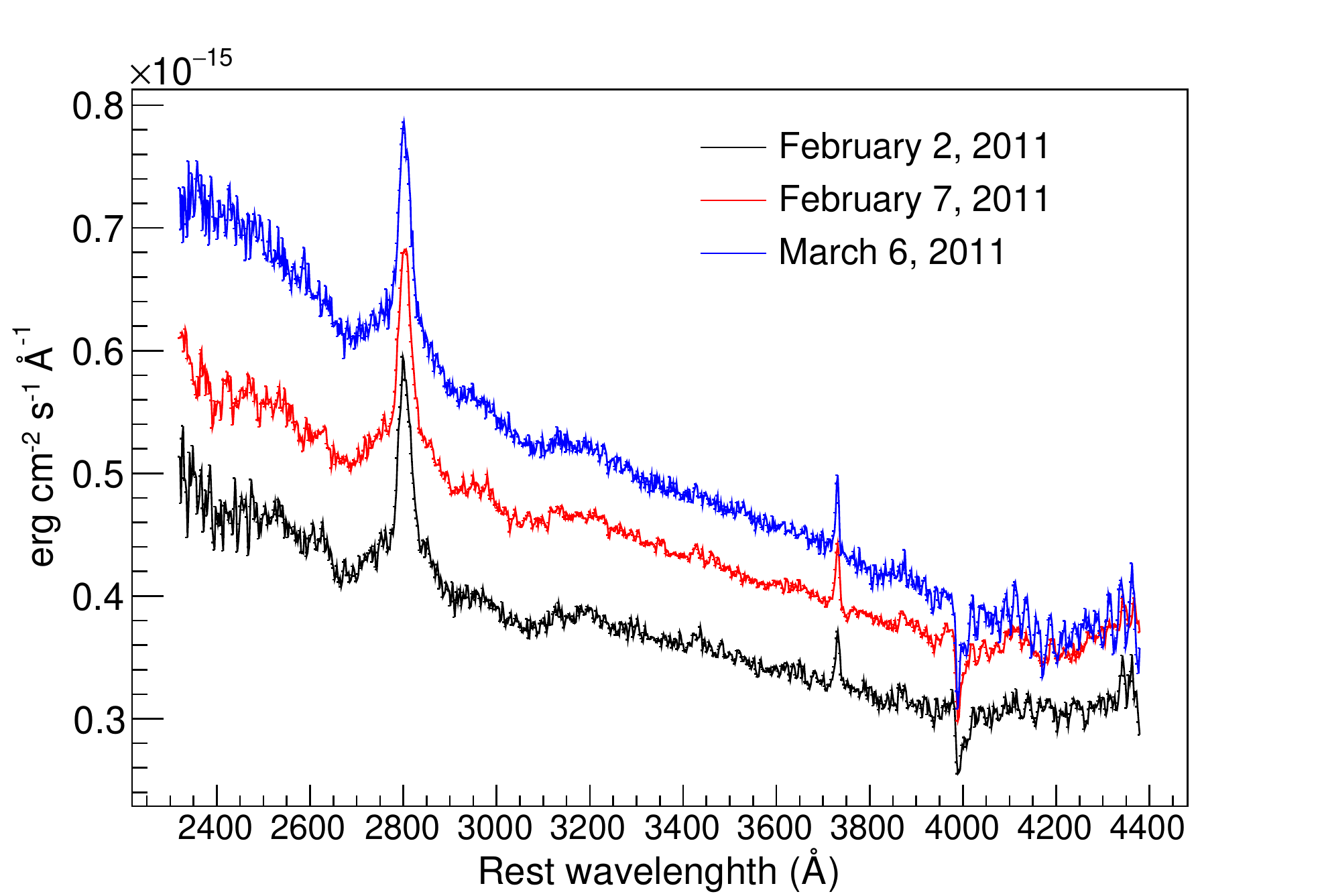}
\caption{SPOL spectra that exhibited the O-II line along with the Mg-II line.}
\label{Mg_O}
\end{figure}

The Mg-II line profile is fitted with the Gaussian function after 
subtracting the underlying  power law continuum which is shown in 
the inset of Figure~\ref{mg}. The Figure~\ref{EW} shows the variation 
of the equivalent width (EW) of the Mg-II line with continuum. It can be seen 
that the EW is inversely proportional to the underlying continuum, which 
suggests that the variations of the Mg-II line are less than that of the continuum. 
It was seen that absolute flux of Mg-II line
was more or less constant with the changes in continuum flux, 
which also means that variations in
Mg-II line are uncorrelated with the variations in continuum flux.
It is also seen that during high continuum flux state the Mg-II line 
is not visible in the optical spectra. It is possible that the line is 
concealed in the continuum during the high flux state. This might be due to 
little or no effect of different levels of the flux state on BLR, from where 
the Mg-II line is thought to be originating. For Ton 599, the optical spectra 
were studied to see the effect of an ionizing continuum on the Mg-II line in 
1981 by \citet{Wills1983}. They reported that the optical spectrum is 
featureless like BL Lac when the source is bright and in the faint state the 
spectrum appeared to be that of normal quasi stellar object (QSO).
Similar results were seen for other blazars CTA 102 \citep{Larionov2016},
3C 454.4 \citep{Raiteri2008} and OJ 248 \citep{Carnerero2015}.
These suggest that enhanced $\gamma$-ray activity of the jet may have 
little or no effect on BLR.

It was observed that, apart from the Mg-II line, some of the optical 
spectra also exhibited a narrow O-II line at the restframe wavelength 
of 3725 {\AA}. These spectra are shown in Figure~\ref{Mg_O}. As 
this line is seen in very few spectra, we did not study the variations of 
this line in detail.

\subsection{The correlation between $\gamma$-ray and radio band}
The correlation between $\gamma$-ray and radio emission is
quantified by using a discrete correlation function (DCF) given by
\citet{Edelson1988}. We use this function on the discrete data set of
$\gamma$-ray and radio band, spanning almost nine years period. These
data sets are linearly detrended \citep{Welsh1999} before estimating DCF.
For all measured pairs ($a_{i}, b_{i}$) having pairwise lag
$\Delta t_{ij} = t_{j} - t_{i}$, unbinned discrete correlation 
function is defined  as,
\begin{equation}
UDCF_{ij} = \frac{(a_{i} - \bar{a})(b_{j} - \bar{b})}{\sqrt{(\sigma_{a}^{2}-e_{a}^{2})(\sigma_{b}^{2}-e_{b}^{2})}},
\end{equation}

DCF is then given by averaging M pairs for which $ (\tau-\Delta\tau/2)\leq \Delta t_{ij}<(\tau+\Delta\tau/2) $,
\begin{equation}
DCF(\tau) = \frac{1}{M}UDCF_{ij},
\end{equation}

The error on DCF is given by,
\begin{equation}
\sigma_{DCF}(\tau) = \frac{1}{M-1} \left\{ \sum_{}^{} [ UDCF_{ij} - DCF(\tau) ] \right\}^{1/2},
\end{equation}

\begin{figure}
\centering
\includegraphics[scale=0.45]{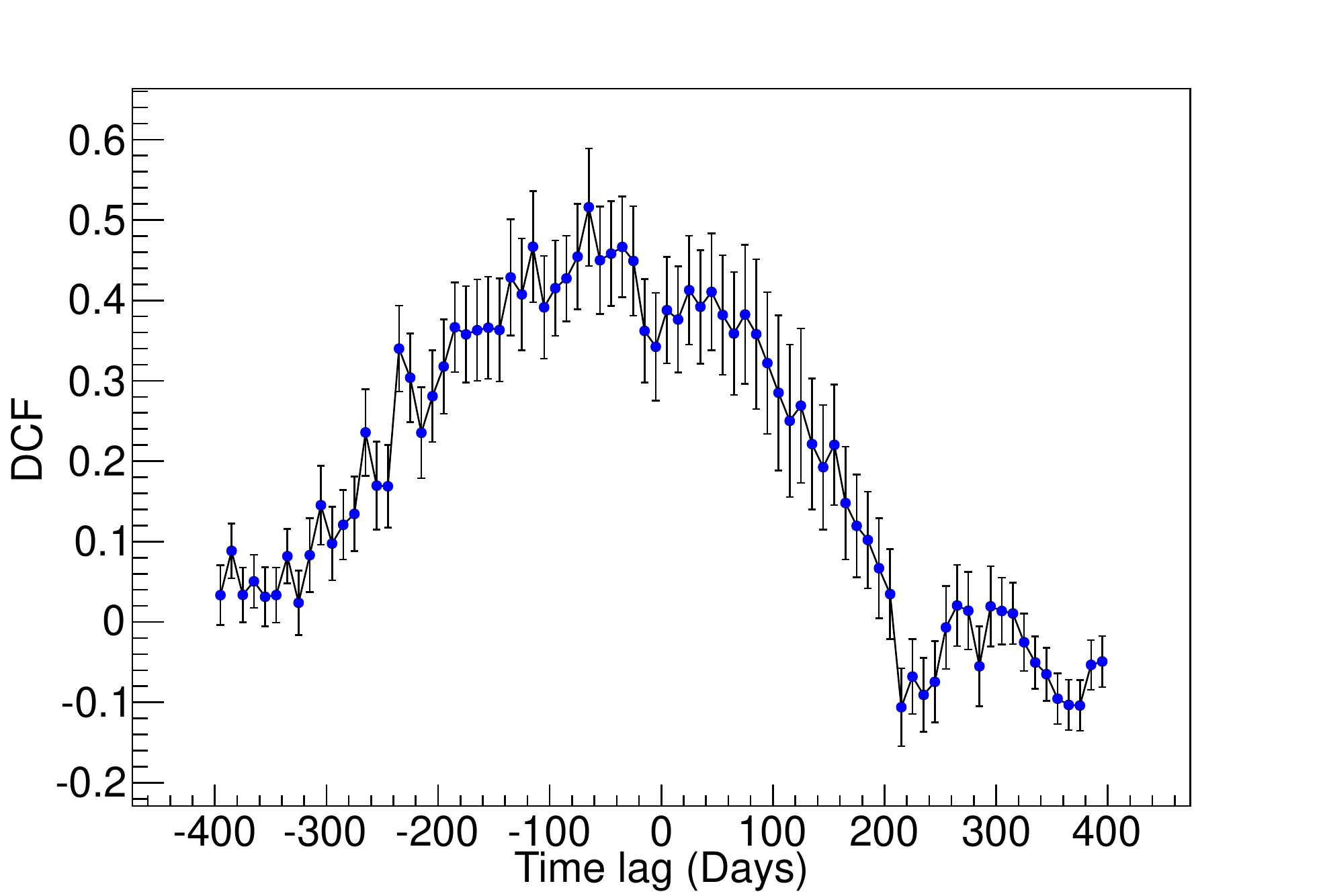}
\caption{DCF between radio and $\gamma$-ray band.
Here negative time-lag means that radio emission lags behind the 
$\gamma$-ray emission.}
\end{figure}

A correlation coefficient of 0.52$\pm$0.07 is seen between the 
radio and $\gamma$-ray band with radio emission lagging behind 
the $\gamma$-ray emission by (65$\pm$10) days.
Similar time-lag was also reported by \citet{Max2014}.
This suggests that $\gamma$-ray emission originates upstream of 
radio emissions. We also checked the DCF by excluding the initial 
three years of radio and $\gamma$-ray data (upto MJD 56000), during 
which the double-peak behavior is observed in radio band. 
It is found that the lag of (65$\pm$10) days is largely coming from 
this period. 
Excluding the data upto MJD 56000, gave slightly
less significant peak at 85.0 day with correlation coefficient of
0.55$\pm$0.17.
Such double-peak structure during this period was also seen at 37 GHz 
(13.7 m telescope at Aalto University Mets$\ddot{a}$hovi Radio Observatory, 
Finland ), 43 GHz (VLBA data) and 230 GHz (1.3mm, the Submillimeter 
Array (SMA)) and the results from these observations are reported by 
\citet{Ramakrishnan2014}. For this source, the radio emission
was lagging the $\gamma$-ray emission by 81.55 days, in the cross
correlation study carried out between 37 GHz 
Mets$\ddot{a}$hovi lightcurves and the $Fermi$-LAT $\gamma$-ray 
lightcurve by \citep{Leon2011}, for the data spanning the period of 2007-2010.


\subsection{The structure function (SF) }
\label{sf}

\begin{figure*}
\centering
\subfigure[SF : Radio band]{
\includegraphics[scale=0.4]{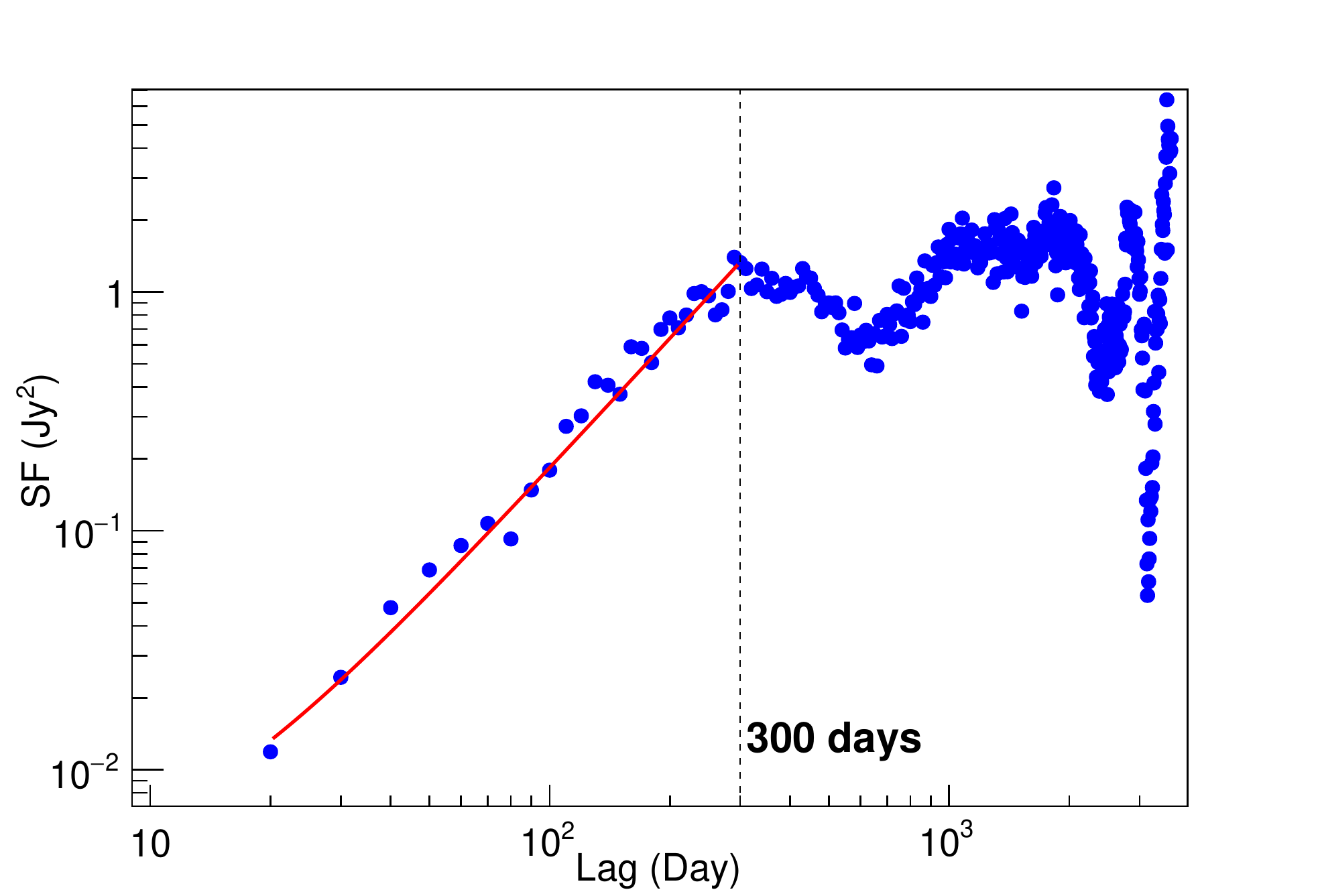}
\label{sf_radio}
}
\subfigure[SF : $\gamma$-ray band]{
\includegraphics[scale=0.4]{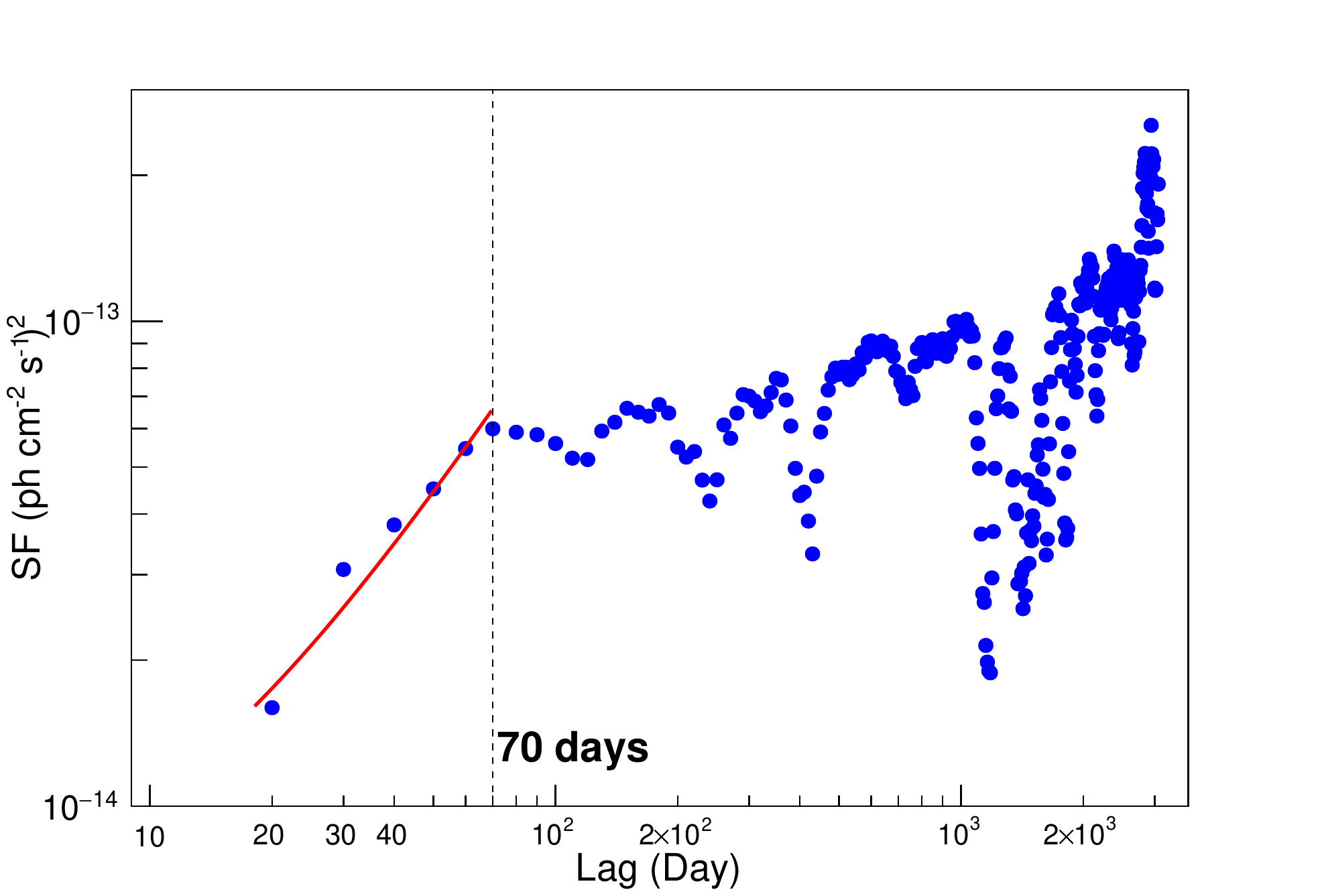}
\label{sf_lat}
}
\caption{The SF for radio
and $\gamma$-ray band. The characteristic timescale is shown by dashed line
in SF plot.}
\end{figure*}

To study the characteristic timescales in the lightcurve, DCF,
Lomb-Scargle periodogram and structure function methods are generally used
in the time series analysis. The periodic variation in radio band 
can be seen in the second panel of Figure~\ref{LC_10}.
To look for any periodicity in this band and also in $\gamma$-ray band,
we used the SF method \citep{Simonetti1985}.
It is used to evaluate the distribution of power in a 
lightcurve. The SF operates in the time domain and  can be
used to examine irregularly sampled lightcurves.
This analysis is done on almost nine years of radio and $\gamma$-ray data.
The first order SF is given by,

\begin{equation}
SF (\tau) = \frac{1}{n} \sum_{ij}  [S(t_i) - S(t_j)]^{2},
\end{equation}

where, S($t_i$) and S($t_j$) is the flux density at
time i and j respectively. Here, $\tau$ is time-lag and n
is number of pairs within the time lag 
$|t_i - t_j|$ 
$\sim \tau$. If there is a variability above the noise level,
the SF rises monotonically with a power law shape and
reaches its first maximum at a plateau. The intersection of the power-law 
fit with the plateau defines the characteristic timescale ($\tau_c$) 
which is an indication of the nature of the process of flux variation 
\citep{Hughes1992}. The SF of Ton~599 at 15 GHz is plotted in 
Figure~\ref{sf_radio}. From the SF analysis of nine years of radio data, 
we find a slope of fitted power-law to be 1.82 and period of $\sim$300 days 
(0.82 year). Similar analysis was done on data collected over $\sim$33 
years from 1980 January to 2012 June by \citet{Liu2014}. They reported 
similar $\tau_c$ of 1.15 $\pm$ 0.05 years at three wavelengths of 4.8, 
8.0 and 14.5 GHz. The work by \citet{Hovatta2007} on the long-term 
variability of AGNs, mentioned this timescale to be 1.21 and 1.36 year,
respectively, at 22 and 37 GHz. We note that, 
at similar radio frequency, 
the value of $\tau_c$ obtained in this work differs by $\sim$28 $\%$, 
with that reported by \citet{Liu2014}.

The SF of $\gamma$-ray lightcurve is shown in Figure~\ref{sf_lat}.
In this band the characteristic timescale and the power-law index  was 
found to be $\sim$70 days and 1.30, respectively. Here, $\tau_c$ in 
$\gamma$-ray band is shorter than that of radio band. This could be an 
indication of slower nature of the mechanism responsible for flux 
variation in radio band as compared to $\gamma$-ray band. In case, 
$\gamma$-ray and radio flux variation are produced by the same mechanism 
(cooling could be possible mechanism), then it could be indicative of a 
decaying nature of such process.
\citet{Marscher2008} suggest that at the base of the jet, explosive
activity injects the surge of energy into the jet. This disturbance 
which appears as a knot of emission, propagates through the acceleration 
and collimation zone. This causes the emission of optical, X-ray and
$\gamma$-ray radiation from the knot, until it exits the zone. 
Owing to synchrotron self-absorption, this zone is opaque at radio 
wavelengths and these are emitted at much downstream of the jet
in the standing shock when the second flare takes place 
\citep{Marscher2008}.

\section{Time-resolved SED}
\label{SED}
Four major peaks can be clearly seen in the $\gamma$-ray lightcurve
shown in Figure~\ref{LC_10} marked as p1, p2, p3, and p4. 
It is also seen that the spectral 
evolution is similar during p1, p2 and p3, however, during p4 it is consistently
harder. In this section, we focus on the spectral studies of these four peaks
which are occurring at four different epochs.

\begin{figure*}
\centering
\includegraphics[scale=0.4]{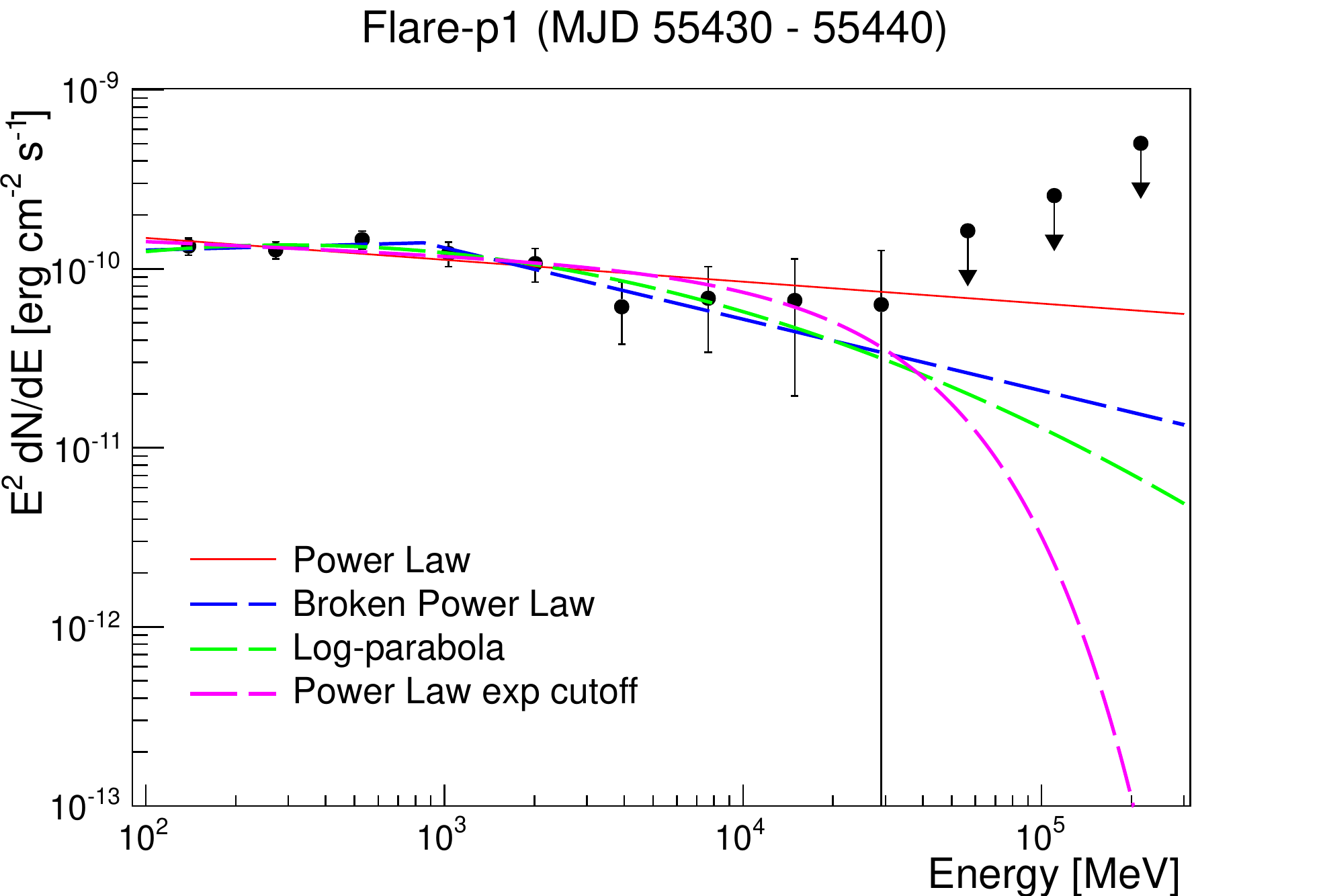}
\includegraphics[scale=0.4]{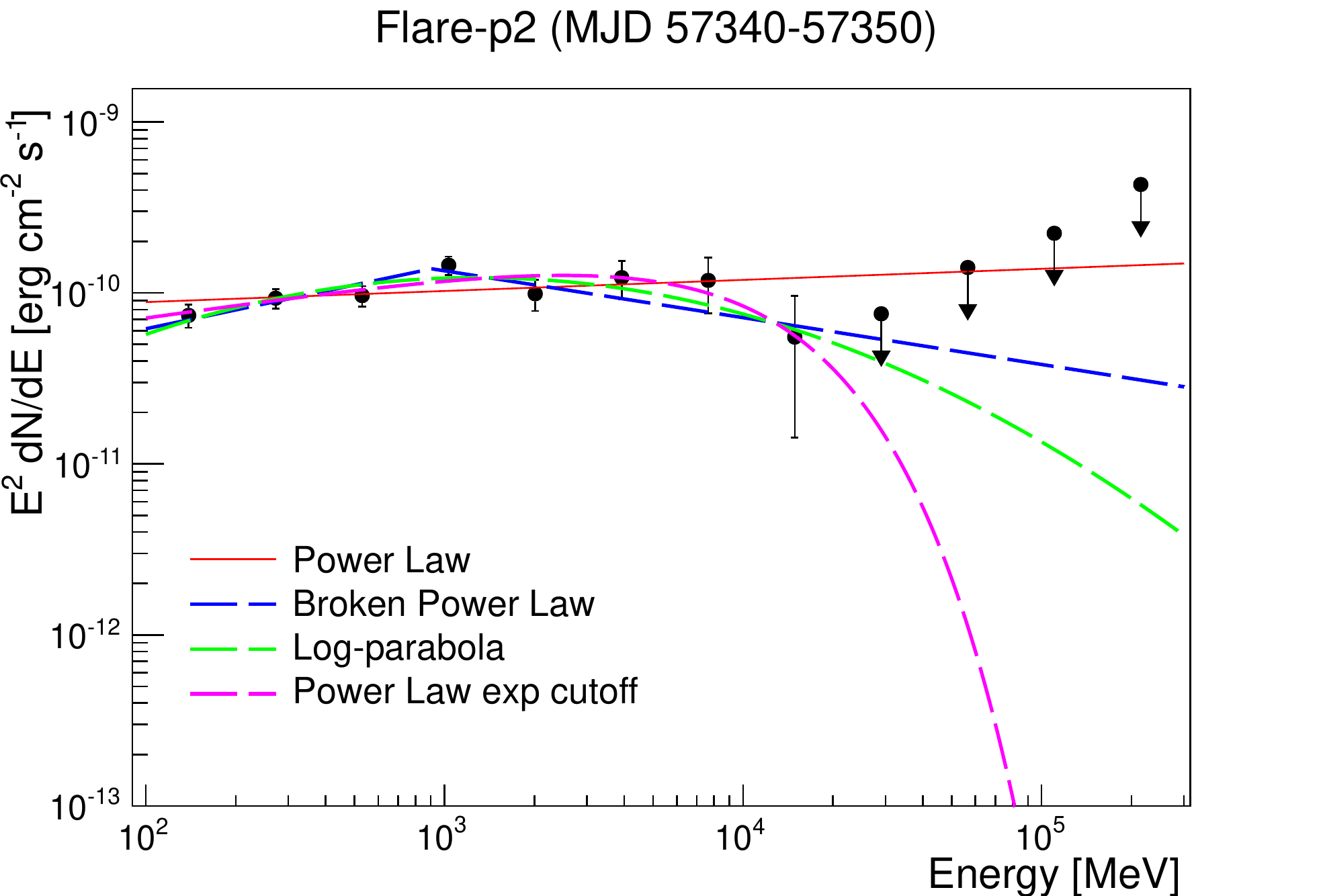}\\
\includegraphics[scale=0.4]{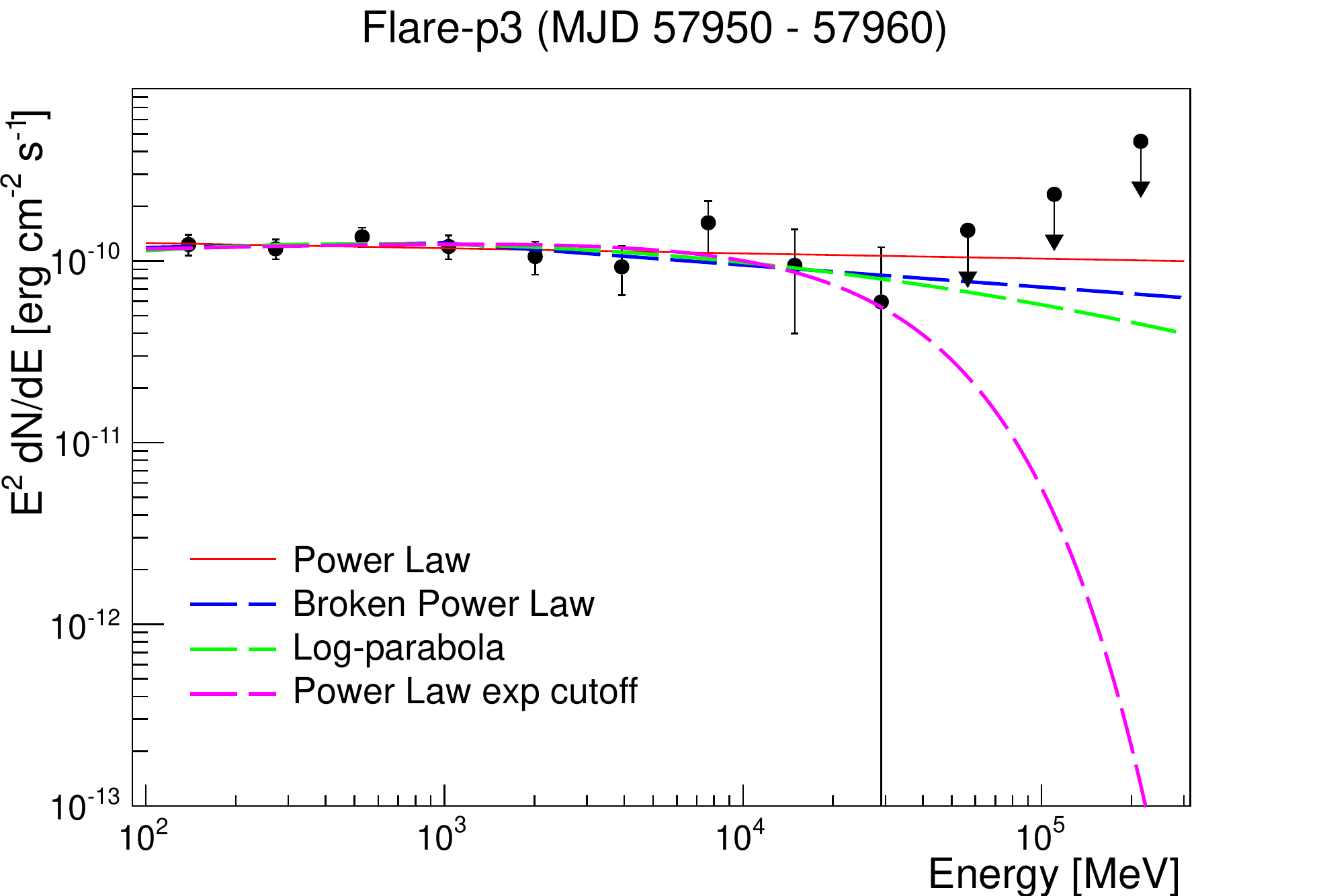}
\includegraphics[scale=0.4]{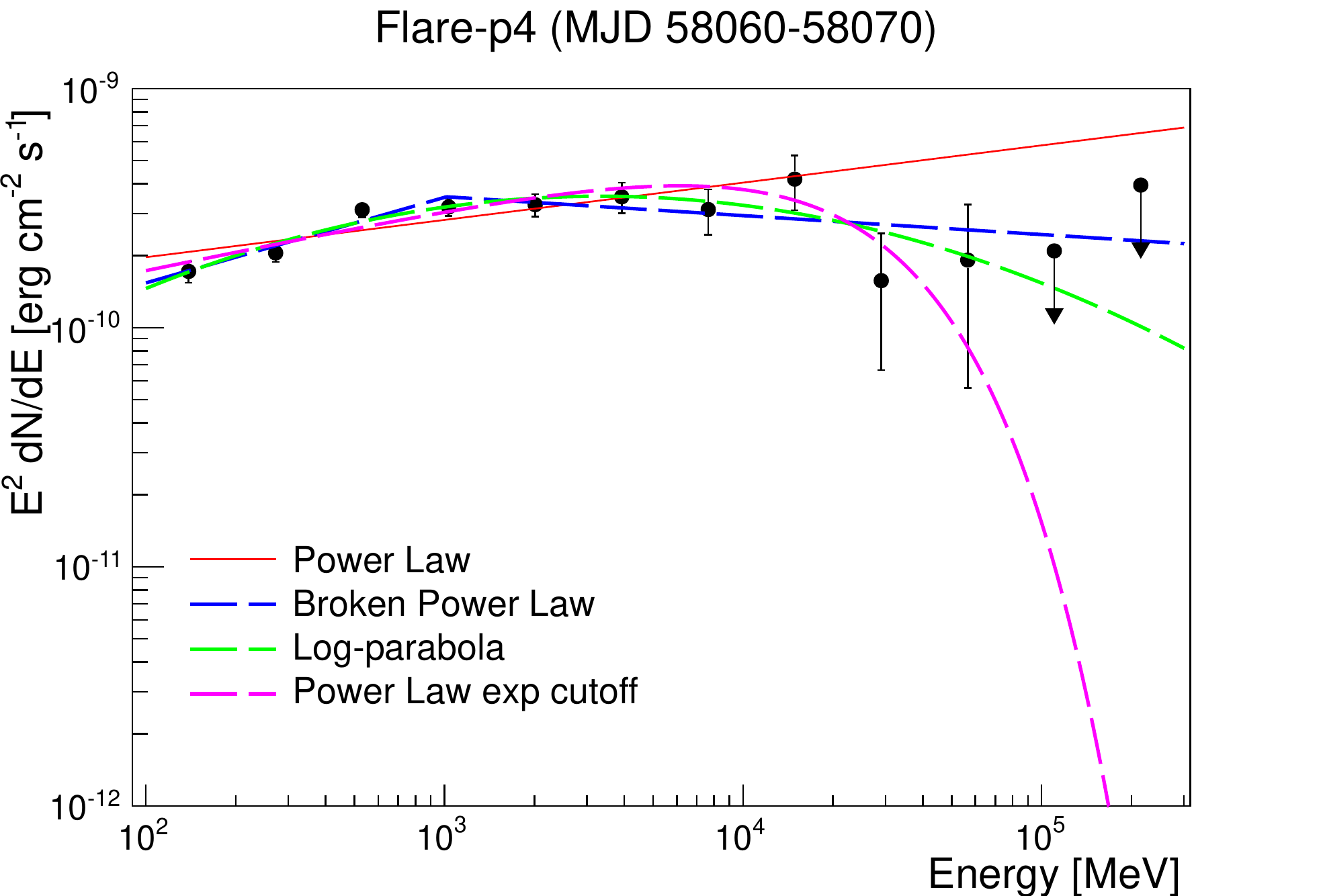}
\caption{The 10-day averaged SED for Ton 599 during four flares p1, p2, p3, and p4, along with best fit PL, BPL, LP and PLEC functions}
\label{sed}
\end{figure*}

In a spectral analysis, SED data points are computed in an equally spaced
logarithmic bins in the energy range of 0.1-300 GeV. Four spectral models
used in the likelihood fit for Ton 599 are defined in 
Equation~\ref{pl}-~\ref{plex}, mentioned in Section~\ref{analysis} 
(respectively as PL, BPL, LP and PLEC). The best-fit parameters of the 
models are mentioned in Table~\ref{SED_para}. 
The fitted models are plotted in Figure~\ref{sed} along with 
SED data points.
The value of log-likelihood ratio test is mentioned in
Table~\ref{SED_para} to check for the improvement of model fit as compared to
power-law model. It can be seen that BPL shows significant improvement in
the fit, compared to PL, at 3 $\sigma$ level for flare p2 and p4, while
LP shows improvement in the fit only for p4.
For other two flares neither BPL, LP nor PLEC give significant
improved fit as compared to PL.
We note that, during p4, 
$\sim$70 $\%$ times the spectral index was harder than that mentioned in 
\textit{Fermi}-LAT 3FGL catalog (see \textit{Panel-3} of Figure~\ref{lc4}),
indicating significant hardening of the spectrum.

\begin{figure}[!h]
\centering
\includegraphics[scale=0.4]{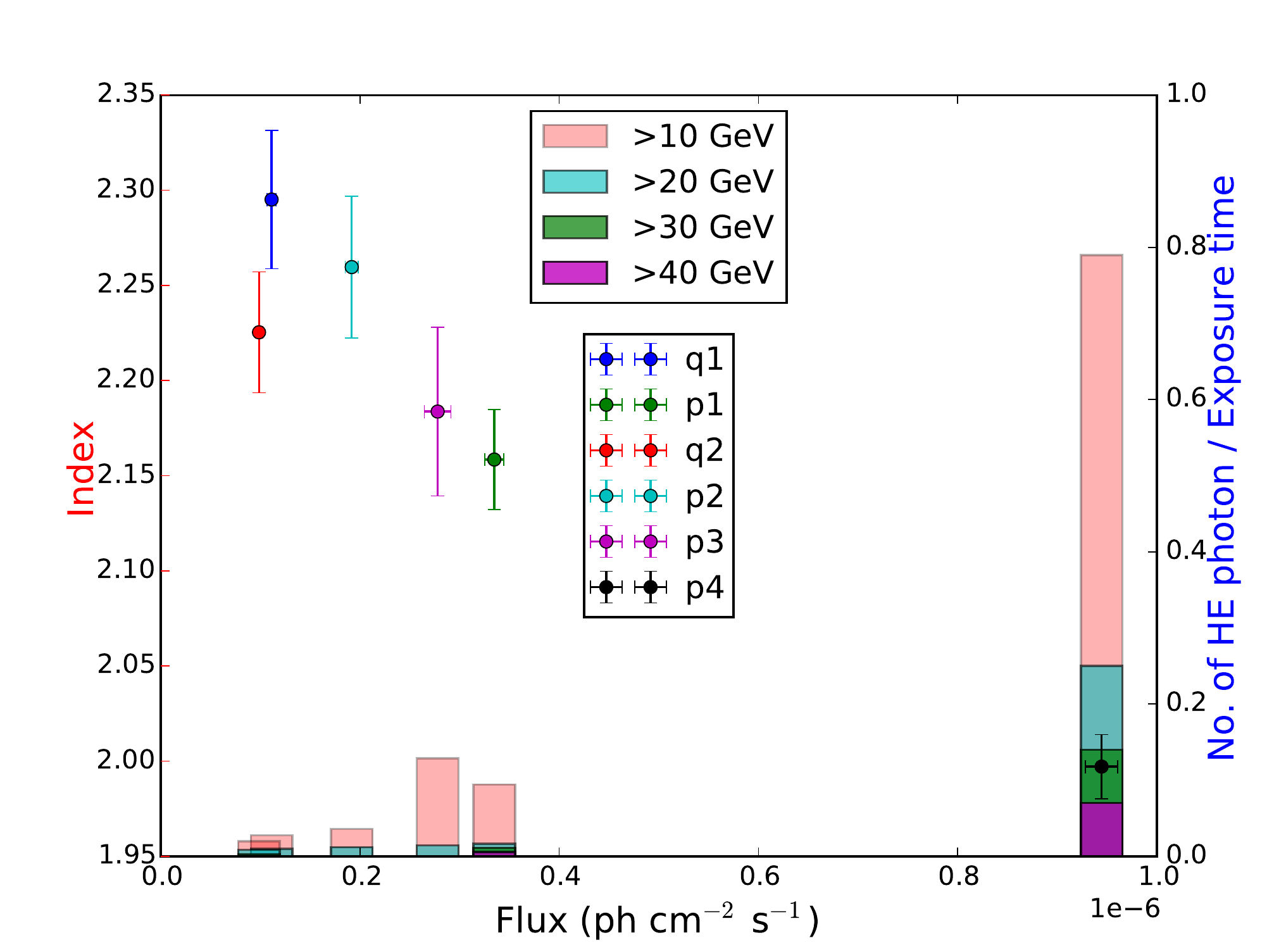}
\caption{A power-law index vs. 0.1-300 GeV flux during six epochs of the
source and corresponding rate of emission of HE photons}
\label{lat_FI}
\end{figure}

\begin{table*}
\centering
\caption{Best fit parameters for PL, BPL, LP and PLEC function, obtained from 
likelihood analysis of spectra during four flares p1, p2, p3, and p4.}
\resizebox{\textwidth}{!}{%
\begin{tabular}{ccccccccc}
\hline
\hline
Flare & Period & Flux &  Luminosity & $\Gamma$ &  &  & Log(likelihood)  & LRT($\chi_{\nu,0.001}^{2}$)\\
PL    & MJD&  10$^{-7}$ph cm$^{-2}$s$^{-1}$ & 10$^{48}$erg s$^{-1}$ &  & & &&\\
\hline
p1 & 55430-55440 & 8.28$\pm$0.51 & 1.57 & 2.12$\pm$0.05 &  &  & -38070.57 & - \\
p2 & 57340-57350 & 5.90$\pm$0.38 & 1.46 & 2.05$\pm$0.05 &  &  & -38123.22 & - \\
p3 & 57950-57960 & 7.62$\pm$0.54 & 1.72 & 2.03$\pm$0.05 &  &  & -39210.28 & - \\
p4 & 58060-58070 & 14.54$\pm$0.60 & 5.58 & 1.84$\pm$0.03 &  & & -50572.40 & - \\
\hline
\hline
& Period & Flux & Luminosity & $\Gamma_1$ & $\Gamma_2$ & Break & Log(likelihood)& \\
BPL    & MJD & 10$^{-7}$ph cm$^{-2}$s$^{-1}$ & 10$^{48}$erg s$^{-1}$ & & & GeV &  \\
\hline
p1 & 55430-55440 & 7.98$\pm$0.53 & 1.37 & 1.96$\pm$0.13 & 2.40$\pm$0.19 & 0.85$\pm$0.62 & -38068.93 & 3.28(6.91)\\
p2 & 57340-57350 & 5.33$\pm$0.87 & 1.63 & 1.86$\pm$0.20 & 2.27$\pm$0.17 & 0.91$\pm$0.39 & -38119.56 & 7.32(6.91)\\
p3 & 57950-57960 & 7.50$\pm$0.58 & 1.62 & 1.98$\pm$0.13 & 2.12$\pm$0.15 & 1.00$\pm$2.19 & -39210.68 & -0.80(6.91)\\
p3 & 58060-58070 & 13.60$\pm$0.63& 4.48 & 1.64$\pm$0.07 & 2.08$\pm$0.07 & 1.02$\pm$0.28 & -50566.64 & 11.52(6.91)\\
\hline
\hline
 & Period & Flux & Luminosity & $\alpha$ & $\beta$ & & Log(likelihood) \\
LP    & MJD& 10$^{-7}$ph cm$^{-2}$s$^{-1}$ & 10$^{48}$erg s$^{-1}$  & & & &&\\
\hline
p1 & 55430-55440 & 7.80$\pm$0.52 & 1.39 & 2.05$\pm$0.07 & 0.06$\pm$0.04 & &-38069.71 & 1.72(10.83)\\
p2 & 57340-57350 & 5.32$\pm$0.45 & 1.27 & 1.75$\pm$0.09 & 0.12$\pm$0.04 & &-38119.57 & 7.30(10.83)\\
p3 & 57950-57960 & 7.43$\pm$0.57 & 1.63 & 1.98$\pm$0.07 & 0.03$\pm$0.03 & &-39209.86 & 0.84(10.83)\\
p4 & 58060-58070 & 13.55$\pm$0.63 & 4.49 & 1.71$\pm$0.05 & 0.07$\pm$0.02 &&-50564.44 & 15.92(10.83)\\
\hline
\hline
& Period & Flux & Luminosity & $\Gamma_{PLEC}$ & E$_c$ & &  Log(likelihood)  \\
PLEC    &  MJD & 10$^{-7}$ph cm$^{-2}$s$^{-1}$ & 10$^{48}$erg s$^{-1}$ & & GeV & & \\
\hline
p1 & 55430-55440 & 8.17$\pm$0.52 & 1.46 & 2.07$\pm$0.08 & 30.19$\pm$40.38 &  & -38111.44 & -81.74(10.83) \\
p2 & 57340-57350 & 5.51$\pm$0.43  &1.30 & 1.74$\pm$0.09 & 9.80$\pm$4.40  &  & -38130.33  & -14.22(10.83)\\
p3 & 57950-57960 & 7.42$\pm$0.55 &1.59  & 1.96$\pm$0.07 & 30.28$\pm$26.27 &  & -39205.93 & 8.70(10.83)\\
p4 & 58060-58070 & 13.98$\pm$0.61 & 4.44 & 1.74$\pm$0.04 & 23.63$\pm$8.27 &  & -50574.64 & -4.48(10.83)\\
\hline
\end{tabular}%
}
\tablecomments{LRT is the value of log-likelihood ratio test, compared with 
the power-law model; $\chi_{\nu,0.001}^{2}$ is the value of reduced $\chi^{2}$
for $\nu$ degree of freedom corresponding to the probability of 0.001}
\label{SED_para}
\end{table*}

The Figure~\ref{lat_FI} shows the correlation between the flux and 
power-law index during four flaring periods and two quiescent states 
of the source (see Figure~\ref{LC_10}). The flux and index are averaged 
over corresponding periods. 
The behavior of hardening of the spectra with the flux can been seen 
in this figure. Also, histograms of high energy photons over total 
exposure periods of corresponding epochs are shown in Figure~\ref{lat_FI}.
It is seen that the during peculiar flare p4, the rate of emission of 
high energy photons has increased, as compared to the previous flares and
quiescent states. 


\section{Discussions}
\label{discussion}

The various jet parameters like Doppler factor, 
jet opening angle, magnetic field in the emission region, size and distance of
emission region from central black hole are estimated for flare p1 and p4
and discussed in this section. For other two flares, p2 and p3, due to lack
of X-ray data, we have not calculated jet parameters during these two
periods.

\subsection{The constraint on Doppler factor}

The minimum Doppler factor can be estimated numerically, by
$\gamma\gamma$ opacity argument and from a detection of the highest energy
photons during the flare \citep{Dondi1995,Ackermann2010}.
Assuming that the optical depth, $\tau_{\gamma\gamma}$($E_{1}$) of the
photon of energy $E_{1}$ to $\gamma\gamma$ interaction is 1, the minimum 
Doppler factor ($\delta_{min}$) can be calculated as,
\begin{equation}
\delta_{min} = \Bigg[ \frac{\sigma_{t} d_{l}^{2} (1+z)^{2} f_{\epsilon} E_{1}}{4t_{var}m_{e}c^{4}} \Bigg]^{1/6},
\end{equation}
where, $\sigma_{t}$ is the Thomson scattering cross section, $E_{1}$ is the 
highest photon energy, $d_{l}$ is the luminosity 
distance which is 4.5 Gpc for Ton 599, $t_{var}$ is the observed variability 
timescale and $f_{\epsilon}$ is the 0.3-8 keV flux (as mentioned in 
Section~\ref{xray}). We have estimated the Doppler factor, assuming 
$t_{var}$ to be the fastest variability timescale (\textit{ln} 2 $\times$ \textit{T}$_{d}$)
observed during each flare. 
Also the size ($R'$) of the $\gamma$-ray emission region is calculated 
from $\delta_{min}$ and $t_{var}$ using $R'$ $\sim$ c$\delta_{min}$ 
$t_{var}$/(1+z). The distance (\textit{d}) of this region from central 
engine is then estimated, assuming $\Gamma$ $\sim$ $\delta_{min}$, as
\textit{d} $\sim$ 2 c$\Gamma^{2} $ $t_{var}$/(1+z) \citep{Abdo2011}.
The estimated value of these jet parameters are given in Table~\ref{jet_para}.
In earlier studies the viewing angle has been estimated to range
from 2$^{\circ}$ to 8$^{\circ}$ \citep{Lahteenmaki1999, Hong2004, 
Hovatta2009,Ramakrishnan2014}, consistent with the value of 5$^{\circ}$.33 
and 4$^{\circ}$.25 (for flare p1 and p4, respectively) found in this work.
The Doppler boosting factor of 28.5 was estimated in earlier work
\citep{Hovatta2009}, which is also consistent with the lower limit of
10.75 and 13.45 (for flare p1 and p4, respectively), estimated in this work.



\subsection{Isotropic luminosity}
The luminosity of the source was calculated for each of the spectral
shapes (PL, BPL, LP and PLEC) at the peak of all four major flares.
The integrated apparent isotropic luminosity is given by,
\begin{equation}
L_{\gamma} = 4\pi d_{l}^{2} \int_{E_{1}}^{E_{2}} E \frac{dN(E)}{dE} dE,
\end{equation}
where, $dN(E)/dE$ is the differential form of spectral model used in the 
analysis. Here, the integral is taken over 100 MeV to the highest energy 
photon detected during each flare. The peak luminosity values during four 
flares are given in the Table~\ref{SED_para}. It is observed that the peak 
luminosity is $\sim$ 3 times higher for p4 than other three flares.
The black hole mass of Ton 599 is found to be (0.79 - 3.47) 
$\times$10$^{8}$ $M_{\odot}$ \citep{Liu2006,Xie2004} and thus the Eddington 
luminosity ($L_{edd}$) is estimated to be (1.0 - 4.37) $\times$ 10$^{46}$ 
erg s$^{-1}$. Now, in order for intrinsic luminosity, 
to be less than Eddington luminosity 
(2(1-cos$\theta$)$L_{\gamma}$ $\leq$ $L_{edd}$), $\theta$ should be in the
range of 5.09-5.65$^\circ$, which is consistent with the results numerically 
estimated from $\gamma\gamma$ opacity argument.

\begin{table*}
\centering
\caption{The calculated jet parameters for flare p1 and p4}
\resizebox{\textwidth}{!}{
\begin{tabular}{ccccccccccc}
\hline
\hline
Flare & 0.3-8 keV flux &  HE photon & $\delta_{min}$  & $\beta_{jet}$  & \textit{d}     & $R'$            & $\theta$ & $f^{sy}_{\epsilon}$ & $\epsilon_{sy}$  & $B_e$\\
      & (10$^{-12}$ erg cm$^{-2}$ s$^{-1}$) & (GeV)     &                 &          & (pc)  & (10$^{15}$cm)& ($^\circ$) & (10$^{-10}$ erg cm$^{-2}$  s$^{-1}$) &  (10$^{-6} $)    & (G)\\
\hline
p1    & 1.32 &  48.02     & 10.75           & 0.9957   & 0.014  &  1.94        &  5.33 & 0.24 & 0.081 & 14.13\\
p4    & 7.67 & 76.94     & 13.45  & 0.9972   & 0.051  &  5.86 & 4.25 & 1.86 & 0.203 & 6.92 \\ 
\hline
\end{tabular}
}
\label{jet_para}
\end{table*}

\subsection{Magnetic field}

\begin{figure}
\centering
\includegraphics[scale=0.45]{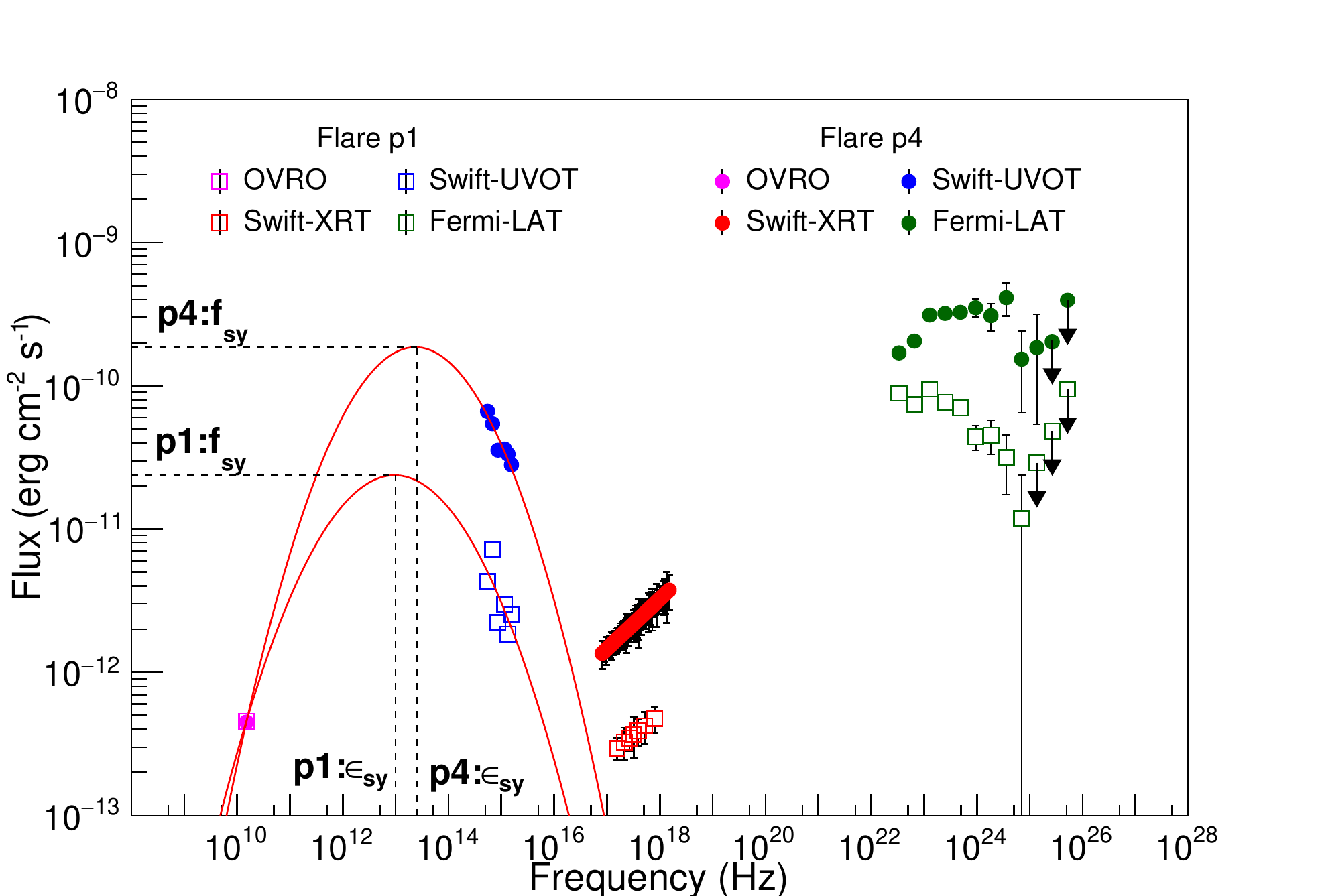}
\caption{SED during p1 and p4, low energy peak fitted with second order polynomial.
Here the radio flux is an upperlimit value and it is considered as flux 
measurement while fitting.}
\label{sed_fit}
\end{figure}

One can also estimate the co-moving magnetic field ($B_e$) in the
emission region from the synchrotron peak flux ($f^{sy}_{\epsilon}$) 
and the synchrotron peak frequency ($\epsilon_{sy}$~=~$h\nu_{sy}/m_e c^{2}$) 
using the equation given by \citet{Bottcher2007} as,

\begin{equation}
B_e = 9 D^{-1}_1 \left[\frac{{d^{4}_{27}} f^{2}_{-10}e^{2}_B}{(1+z)^{4} \epsilon_{sy,-6} R'^{6}_{15} (p-2)} \right]^{1/7} G,
\end{equation}

where, $f_{-10}~=~f^{sy}_{\epsilon}$/10$^{-10}$ erg cm$^{-2}$ s$^{-1}$, 
$\epsilon_{sy,-6}~=~\epsilon_{sy}$/10$^{-6}$, $R'_{15}= R'/$10$^{15}$ cm, 
$e_B$ is the equipartition parameter in the co-moving frame, 
$d_{27} = d_l/$10$^{27}$ cm and $p$ is the spectral index of underlying
electron spectrum. $D_1 = D/10$, where $D$ is the Doppler factor.
For a better estimate of synchrotron peak, we fitted the second order 
polynomial to the lower-energy part of the SED
for flare p1 and p4. These fits are shown in Figure~\ref{sed_fit}.
The values of $f_{sy}$ and $\epsilon_{sy}$ obtained from these fits
are mentioned in Table~\ref{jet_para}.
We have adopted an underlying electron spectral index, 
$p=3.4$ \citep{Celotti2008} for both the flares.
Assuming the equipartition of energy, magnetic field in the emitting region 
during each flare is estimated, using corresponding estimated values of 
Doppler factor ($\delta_{min}$), $R'$, $f_{sy,-10}$ and $\epsilon_{sy,-6}$. 
These are given in the last column of Table~\ref{jet_para}. 
The $B_e$ estimated here is 14.13~G and 6.92~G for p1 and p4, 
respectively. The magnetic field of few Gauss generally is found 
by the modeling of FSRQs. For Ton~599, 
it was found to be 4 G \citep{Ghisellini2010} and 5 G \citep{Celotti2008}.
This $B_e$, of the order of a few Gauss would produce much lower
break ($\gamma_b$ $\sim$ 137) in an injected particle distribution for the 
emission region of size $\sim$10$^{15}$ cm, which is obtained in this work.
This suggests that radiative losses would be much more stronger in the 
emitting region. We also observed the X-ray spectrum to be harder ($\sim$1.7), 
which can be associated with EC mechanism for FSRQs, because SSC component
tends to be softer \citep{Celotti2008}. Looking at the composition of the
jet, \citet{Ghisellini2010} reported that the jets of FSRQs are matter 
dominated i.e much of the jet power goes into forming large radio lobe 
structures and it transforms only few per cent of their kinetic energy into 
radiation.

\begin{figure}
\centering
\includegraphics[scale=0.45]{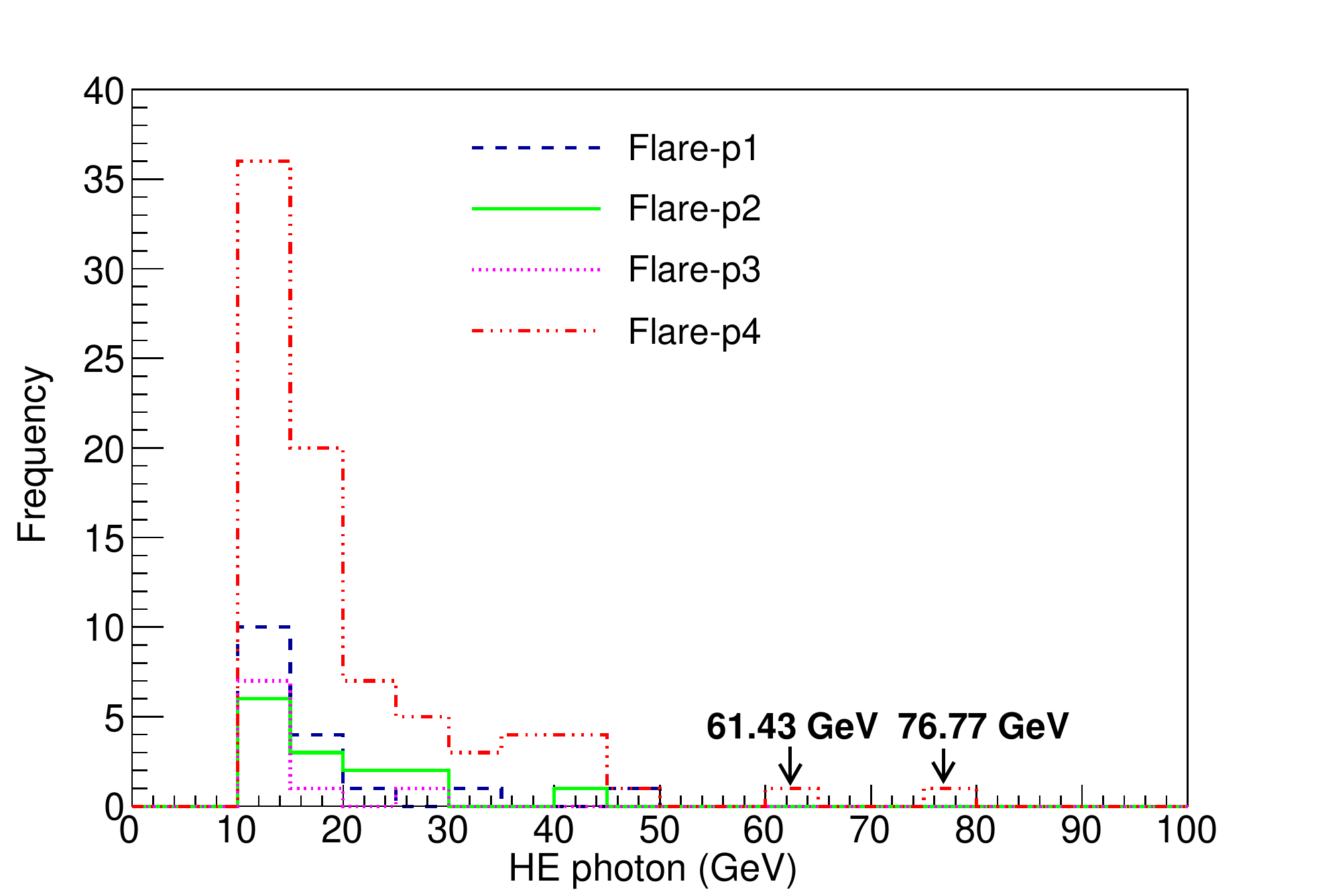}
\caption{The distributions of HE $\gamma$-ray photons emitted from the 
region of 0.5$^\circ$ radius around the source position during 
flare-p1,p2,p3 and p4.}
\label{HE_photon}
\end{figure}

\subsection{Size of BLR}

The BLR is assumed to be a thin spherical shell and its distance
from central black hole is given by \citet{Ghisellini2010} as,
\begin{equation}
R_{BLR} = 10^{17} L^{1/2}_{d,45} cm,
\label{blr}
\end{equation}
where, R$_{BLR}$ is the distance of BLR from the central engine and
$L_{d,45}$ is the accretion disk luminosity in the units of
10$^{45}$ erg s$^{-1}$. Assuming $L_{BLR}$ to be 10$\%$ of $L_{d}$, as
this is the fraction of disk luminosity ($L_{d}$) re-emitted by the broad lines,
$L_{d}$ is estimated to be in the range of (4.47-6.00) $\times$ 10$^{45}$
erg s$^{-1}$ \citep{Shaw2012,Xiong2014,Ghisellini2010}. Hence,
$R_{BLR}$ is estimated to be (2.11-2.45) $\times$ 10$^{17}$ cm (0.068-0.079 pc).

\subsection{Rise and decay time distribution and change in jet parameters over the $Fermi$-LAT era}
\begin{figure}
\centering
\includegraphics[scale=0.45]{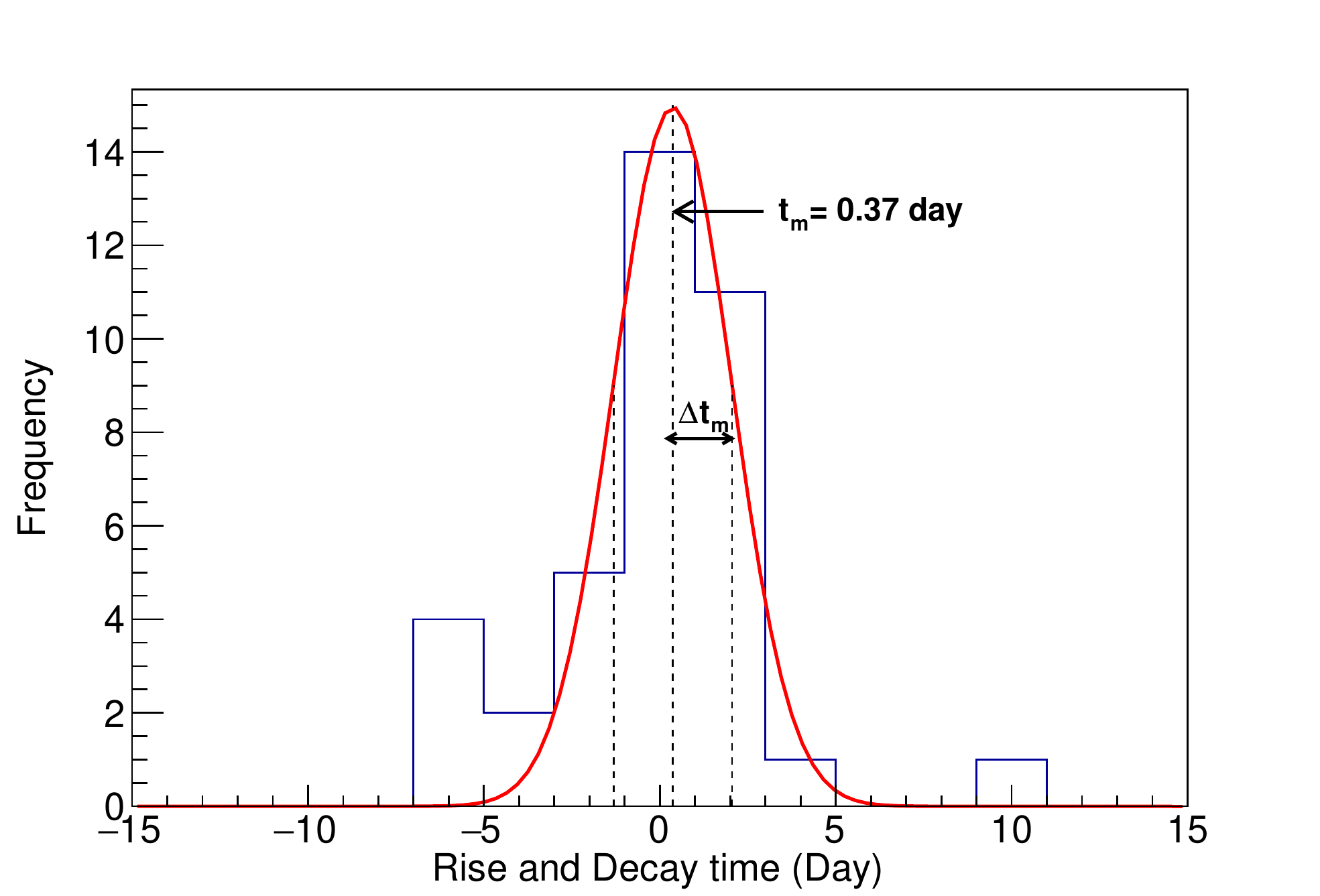}
\caption{The distribution of rise and decay times that are obtained from
flare fitting. The positive time corresponds to rise time and negative time corresponds to decay time}
\label{figTimeDist}
\end{figure}

The excellent sampling of flux measurements in the high energy $\gamma$-ray
band from $Fermi$-LAT have enabled us to scan the sources in shorter timescales.
Figure~\ref{figTimeDist} shows the distribution of all rise and decay times 
obtained using flare fitting method. It can be seen that the distribution peaks 
($t_m$) at 0.37~day with standard deviation ($\Delta t_m$) of 1.68 days. 
Considering variability time range from  $t_m-\Delta t_m $ to  $t_m+\Delta t_m $,
the Doppler factor varies by $\sim$33 $\%$ and $\sim$23 $\%$ 
for flare p1 and p4, respectively.

\section{Conclusion}
\label{Conclusion}

1. The jet parameters and emission of HE photons: The $\gamma$-ray emission region is found to be 
located between (0.024-0.051) pc from the central engine during flare p1 and p4. 
The flare-p4 is the highest ever detected flaring state in $\gamma$-ray 
band for this source. We also note that during this period more than 
50 photons with the energy above 10 GeV were detected in the span 
of $\sim$40 days. The two highest energy photons, that were observed, 
had the energies of 76.9 GeV and 61.9 GeV. 
This is possible if the emission region lies near outer edge 
or outside the BLR. For emission region to be near the outer edge of BLR 
(i.e. 0.08 pc), Doppler factor would have increased upto $\sim$35, during p4,
implying the magnetic field of $\sim$3 G. 
The distributions of HE $\gamma$-ray photons during 
four flares are shown in Figure~\ref{HE_photon}. Similar behavior
was seen in 3C~454.3 outburst in 2014 May-June with detection of 
several HE photons above 20 GeV, including one at 45 GeV \citep{Britto2016}.

Alternatively, the phenomenon of emission of bunch of HE 
photons during flare-p4 for this source, can be explained if the emission 
region is moving on helical path of the jet \citep[for helical jet models, see, 
for example,][]{Steffen1995, Villata1999, Raiteri2017}. 
In this case, it is possible that the up-scattered HE $\gamma$-ray 
photons from the emission region may not interact with the BLR clouds as they 
will not come in the path between the emission region and the observer.
The helical jet model proposed by \citet{Steffen1995}, was used in
the study of 1980-2002 VLBI observations of Ton 599 \citep{Hong2004}.
For CTA 102, a helical jet model was used by \citet{Shukla2018} to explain the
short-timescale variability on less than the light travel-time across the 
black hole which is $\sim$70 minutes. 

2. The $\gamma$-ray and radio emission connection : The $\gamma$-ray and radio 
emissions are closely correlated due to the fact
that the emission in both these bands are beamed and that both the emissions
are produced co-spatially in the jet \citep[for example,][]{Leon2011, Richards2014, Fan2016}.
The decomposition of the major flare reveals several asymmetric sub-flaring
structures in the $\gamma$-ray lightcurves. This might be due to an
ejection of new particles or an acceleration of relativistic particles
by passing of initial phase of a shock wave during the rising part of the
flare and successive settling down of turbulence during decaying stage.
The same shock then might be producing the radio emission, which is
peaking $\sim$60 days later. For $\gamma$-ray flare of OJ~287, it was
proposed that the flare is triggered by the interaction of moving blobs
of plasma and the standing shock \citep{Agudo2011}.
\citet{Leon2012} observed the high spectral turnover frequencies in radio band.
They suggested that such observations reveal the presence of emerging
disturbances in the jet which are more likely to be responsible for 
the high flux states of the $\gamma$-ray emission.

\acknowledgements

We thank referee for her/his constructive comments which has 
significantly improved the manuscript. 
This work has made use of $\textit{Fermi}$ data, obtained from Fermi 
Science Support Center,
provided by NASA's Goddard Space Flight Center (GSFC).
The data, software and web tools obtained from
NASA's High Energy Astrophysics Science Archive Research Center (HEASARC),
a service of GSFC are used. The XRT Data Analysis Software (XRTDAS) developed under 
the responsibility of the ASI Science Data Center (ASDC), Italy has been used in this
work. Also, data from the Steward Observatory spectropolarimetric
monitoring project are used. This program is supported by Fermi Guest Investigator
grants NNX08AW56G, NNX09AU10G, NNX12AO93G, and NNX15AU81G. In this research we
used data from the OVRO 40-m monitoring program \citep{Richards2011} which is supported
in part by NASA grants NNX08AW31G, NNX11A043G, and NNX14AQ89G and NSF grants
AST-0808050 and AST-1109911.
We used Enrico, a community-developed Python package to simplify
Fermi-LAT analysis \citep{enrico}.

\bibliography{Ton599_18Aug18}
\bibliographystyle{aasjournal}

\end{document}